\documentclass[sigconf]{acmart}
\usepackage{amsmath,amsfonts}
\usepackage[ruled,vlined]{algorithm2e}
\usepackage{tabularx}
\usepackage{subfig}
\usepackage{graphicx}
\usepackage{physics}
\usepackage{textcomp}
\usepackage{fancyhdr}
\usepackage{enumitem}
\usepackage{afterpage}
\usepackage{array}
\usepackage{bm}
\usepackage{float}
\usepackage[normalem]{ulem}
\usepackage{caption}
\usepackage{environ}
\usepackage{makecell}
\usepackage{textcomp}
\usepackage{booktabs}
\usepackage{balance}
\usepackage{pifont}% http://ctan.org/pkg/pifont
\newcommand{\cmark}{\ding{52}}%
\newcommand{\xmark}{\ding{56}}%
\newcolumntype{P}[1]{>{\centering\arraybackslash}p{#1}}
\colorlet{shadecolor}{gray!20}

\def\BibTeX{{\rm B\kern-.05em{\sc i\kern-.025em b}\kern-.08em
    T\kern-.1667em\lower.7ex\hbox{E}\kern-.125emX}}

\newcommand{\revision}[1]{\textcolor{black}{#1}}
\newcommand{\sol}{\textsc{Kairos}}
\newcommand{\solplus}{\textsc{Kairos+}}

\newcommand{\drs}{DRS}
\newcommand{\clk}{CLKWRK}
\newcommand{\ribbon}{Ribbon}

\copyrightyear{2023}
\acmYear{2023}
\setcopyright{acmlicensed}
\acmConference[HPDC '23] {Proceedings of the 32nd International Symposium on High-Performance Parallel and Distributed Computing}{June 16--23, 2023}{Orlando, FL, USA.}
\acmBooktitle{Proceedings of the 32nd International Symposium on High-Performance Parallel and Distributed Computing (HPDC '23), June 16--23, 2023, Orlando, FL, USA}
\acmPrice{15.00}
\acmISBN{979-8-4007-0155-9/23/06}
\acmDOI{10.1145/3588195.3592997}

\renewcommand\footnotetextcopyrightpermission[1]{}

\newtheorem{definition}{Definition}

\begin{document}

% \title{\sol{}: Exploiting Hardware Heterogeneity for Machine Learning Model Inference Serving}
\title{\sol{}: Building Cost-Efficient Machine Learning Inference Systems with Heterogeneous Cloud Resources}

\author{Baolin Li}
\affiliation{%
  \institution{Northeastern University}
  \country{}}
%   \city{Boston}
%   \state{MA}}
% \email{li.baol@northeastern.edu}

\author{Siddharth Samsi}
\affiliation{%
  \institution{MIT}
  \country{}}
%   \city{Lexington}
%   \state{MA}}
% \email{karen.gettings@ll.mit.edu}

\author{Vijay Gadepally}
\affiliation{%
  \institution{MIT}
  \country{}}
%   \city{Lexington}
%   \state{MA}}
% \email{vijayg@ll.mit.edu}

\author{Devesh Tiwari}
\affiliation{%
  \institution{Northeastern University}
  \country{}}

\renewcommand{\shortauthors}{B. Li et al.}

%%
%% The abstract is a short summary of the work to be presented in the
%% article.
\begin{abstract}
Online inference is becoming a key service product for many businesses, deployed in cloud platforms to meet customer demands. Despite their revenue-generation capability, these services need to operate under tight Quality-of-Service (QoS) and cost budget constraints. This paper introduces \sol{}~\footnote{\sol{} has been accepted at the 32nd ACM International Symposium on High-Performance Parallel and Distributed Computing (HPDC '23)}, a novel runtime framework that maximizes the query throughput while meeting QoS target and a cost budget. \sol{} designs and implements novel techniques to build a pool of heterogeneous compute hardware without online exploration overhead, and distribute inference queries optimally at runtime. Our evaluation using industry-grade machine learning (ML) models shows that \sol{} yields up to 2$\times$ the throughput of an optimal homogeneous solution, and outperforms state-of-the-art schemes by up to 70\%, despite advantageous implementations of the competing schemes to ignore their exploration overhead.
\end{abstract}
\settopmatter{printacmref=false}

% \begin{CCSXML}
% <ccs2012>
% <concept>
% <concept_id>10010520.10010521.10010537.10003100</concept_id>
% <concept_desc>Computer systems organization~Cloud computing</concept_desc>
% <concept_significance>500</concept_significance>
% </concept>
% <concept>
% <concept_id>10010147.10010257</concept_id>
% <concept_desc>Computing methodologies~Machine learning</concept_desc>
% <concept_significance>300</concept_significance>
% </concept>
% <concept>
% <concept_id>10010520.10010521.10010542.10010546</concept_id>
% <concept_desc>Computer systems organization~Heterogeneous (hybrid) systems</concept_desc>
% <concept_significance>500</concept_significance>
% </concept>
% </ccs2012>
% \end{CCSXML}

% \ccsdesc[500]{Computer systems organization~Cloud computing}
% \ccsdesc[300]{Computing methodologies~Machine learning}
% \ccsdesc[500]{Computer systems organization~Heterogeneous (hybrid) systems}


\maketitle
\pagestyle{empty}

\vspace{-1mm}
\section{Introduction}
\label{sec:intro}

As machine learning (ML) models are becoming widely adopted in commercial services, the service providers will utilize cloud computing resources to serve their customers, and online inference has become a highly critical application for both on-premise and public cloud computing platforms~\cite{zhang2019mark,gupta2020architectural,hazelwood2018applied}. As a result, an increasing amount of research effort is dedicated to improving the capability of cloud systems for inference workloads~\cite{crankshaw2017clipper,crankshaw2020inferline,ning2019deep,choi2021lazy,ali2020batch,wang2021morphling}. Serving ML inference is particularly challenging because they pose additional constraints and objectives beyond meeting latency deadlines. For example, business service providers can utilize the pay-as-you-go model to rent cloud computing instances, but they seek the following desirable objectives: (1) meet the quality-of-service target (QoS constraint, e.g., 99\% of queries finish within 100ms); (2) efficient under a fixed cost budget; (3) process as many queries as possible per time unit (i.e., high query throughput). 

Cloud platforms provide a wide range of virtual machines (VMs), and each comes with different hardware types (e.g., different CPU, GPU, and memory). While there have been previous attempts at providing partial solutions to exploit hardware heterogeneity in datacenter~\cite{delimitrou2013paragon,tumanov2016tetrisched,banerjee2020inductive}, edge~\cite{zhou2018s,xiang2019pipelined}, and cloud~\cite{hu2021scrooge,li2021ribbon,gupta2020deeprecsys}, we lack a complete solution to achieve all the desirable properties (Sec.~\ref{sec:related_work}). In particular, prior schemes do not consider the full aspects of inference serving: \textit{heterogeneous resource allocation and intelligent query distribution among allocated hardware resources.}

Note that a heterogeneous pool of cloud compute instances (a mixture of GPUs and CPUs) appear naturally more promising for inference serving as they provide the opportunity to balance the trade-off between cost and performance (QoS target). More powerful and expensive instances can be used toward satisfying strict QoS targets for larger queries. Less powerful and relatively less expensive instances can be used for executing smaller queries that will not violate their QoS on such instances, and thereby, provide a chance to reduce the overall cost of the query serving system. Consequently, many prior techniques have opportunistically taken advantage of hardware heterogeneity to improve query throughput or meet QoS target~\cite{delimitrou2013paragon,tumanov2016tetrisched,zhou2018s,xiang2019pipelined,gupta2020deeprecsys}. However, none of them provide a systematic methodology to efficiently optimize the heterogeneous configuration (i.e., determine the number of GPUs and CPUs of different types). \textbf{Therefore, while prior works are heterogeneity-aware, they do not proactively optimize the hardware heterogeneity under a cost budget.}

\begin{table*}[!t]
\centering
\caption{Overview of related works and \sol{}.}
\vspace{-0.1cm}
\scalebox{0.85}{
\begin{tabular}{cP{1.3cm}P{1.3cm}P{1.2cm}P{1.6cm}P{2.4cm}P{2.cm}P{5.4cm}} 
%  \hline
%  \multicolumn{1}{|c|}{\textbf{Category}} & \multicolumn{1}{|c|}{\textbf{Instance types}} & \multicolumn{1}{|c|}{\textbf{Size}} \\ [0.5ex] 
\toprule
\textbf{ } & \textbf{Inference QoS} & \textbf{Through-  put} & \textbf{Cost} & \textbf{Query Mapping} & \textbf{Proactive in Heterogeneity} & \textbf{No Online Exploration} & \textbf{Miscellaneous Notes} \\ 
\midrule
\midrule
Paragon~\cite{delimitrou2013paragon} & \xmark & \cmark & \xmark & \cmark & \xmark & \xmark & Requires prior data for training \\
\midrule
TetriSched~\cite{tumanov2016tetrisched} & \xmark & \xmark & \xmark & \cmark & \xmark & \cmark & Supports user-based reservation\\  
\midrule
S$^3$DNN~\cite{zhou2018s} & \cmark & \cmark & \xmark & \cmark & \xmark & \cmark & Uses supervised CUDA stream\\  
\midrule
DART~\cite{xiang2019pipelined} & \cmark & \cmark & \xmark & \cmark & \xmark & \xmark & Profiles layers and applies parallelism\\
% \midrule
% Clockwork~\cite{gujarati2020serving} & \cmark & \cmark & \xmark & \xmark & \xmark & \cmark & Consolidates latency for predictability \\
\midrule
Scrooge~\cite{hu2021scrooge} & \cmark & \cmark & \cmark & \xmark & \xmark & \xmark & Chain execution of media applications \\
\midrule
Ribbon~\cite{li2021ribbon} & \cmark & \cmark & \cmark & \xmark & \cmark & \xmark & Bayesian Optimization for allocation \\
\midrule
DeepRecSys~\cite{gupta2020deeprecsys} & \cmark & \cmark & \xmark & \cmark & \xmark & \xmark & Schedules using profiled threshold \\ 
\midrule
Clockwork~\cite{gujarati2020serving} & \cmark & \cmark & \xmark & \cmark & \xmark & \cmark & Consolidates latency for predictability \\
\midrule
\sol{} & \cmark & \cmark & \cmark & \cmark & \cmark & \cmark & Full heterogeneity support\\

\bottomrule
\end{tabular}}
\label{table:intro}
% \vspace{-0.3cm}
\end{table*}

In fact, we show that some heterogeneous configurations can perform significantly worse than an otherwise cost- and QoS-equivalent homogeneous configuration (Sec.~\ref{sec:motiv}). Determining a heterogeneous configuration requires online evaluation of multiple potential candidates. Unfortunately, this approach is not suitable when the query load changes or other system parameters change, since it requires invoking the exploration process frequently and potentially evaluating configurations that are worse than homogeneous configurations. This has been the main hindrance for the community to exploit heterogeneous computing hardware. \textbf{\sol{} breaks this limitation and designs novel techniques to take full advantage of hardware heterogeneity while meeting QoS constraints under a given cost budget.}\newline

% \vspace{2mm}
\noindent\textbf{Summary of Contributions.} \textit{We design and implement \sol{}, a novel runtime framework to maximize throughput under cost budget and QoS constraints for machine learning inference tasks.} \sol{} breaks away from searching the complex and vast configuration space of heterogeneous hardware. Instead, \sol{} devises two techniques to quickly find a high-throughput heterogeneous configuration without exploring.

First, \sol{} designs an efficient query-distribution mechanism to distribute queries among different cloud computing instances for any given heterogeneous configuration to maximize throughput -- formulating this as a bipartite matching problem and solving it efficiently. Second, \sol{} approximates the upper bound of the throughput that a heterogeneous configuration can provide at the best. Then, \sol{} uses the similarity in top-ranked heterogeneous configurations to pick the most promising heterogeneous configuration without online evaluation. Our evaluation confirms that \sol{}'s configuration choice is often the near-optimal configuration across different machine learning models in production, where the optimal configuration is determined via exhaustive offline search of all heterogeneous configurations. 

We have leveraged industry-grade deep learning models to drive the evaluation of \sol{}'s effectiveness~\cite{gupta2020deeprecsys} -- although we note that \sol{}'s design is generic and not tuned for particular kinds of ML models. Our evaluation shows that compared to the optimal homogeneous configuration, \sol{} is able to significantly increase the throughput (by up to 2$\times$) under the same QoS target and cost budget. \sol{} outperforms the state-of-the-art schemes in this area (Ribbon, DeepRecSys, and Clockwork~\cite{li2021ribbon,gupta2020deeprecsys,gujarati2020serving}) by up to 70\%, despite advantageous implementations of those competing schemes by ignoring the exploration overheads and improving the query distribution technique. Our proposed solution, \sol{}, is publicly available as an open-source package at \textit{\url{https://doi.org/10.5281/zenodo.7888058}}.

\section{Related Work}
\label{sec:related_work}

% \todo{Lacks some reference to scheduling algorithms}

Table~\ref{table:intro} lists the relevant works in exploiting heterogeneous hardware and inference serving. Overall, \sol{} is the only work that satisfies all the desirable properties (table header from left to right): (i) meets QoS for inference queries; (ii) has service throughput requirement; (iii) is aware of heterogeneous hardware cost; (iv) intelligently distributes (or maps) queries among resources; (v) proactively allocates and optimizes heterogeneous resources; and (vi) does not need prior knowledge to train a model or perform online exploration. While some previous works are heterogeneity-aware (i.e., can efficiently use available heterogeneous hardware), they do not proactively configure the heterogeneity to optimize other aspects: query throughput, QoS, and cost budget. 

Latency-critical applications are commonly studied in large-scale datacenter and cloud systems~\cite{patel2020clite,mars2011bubble,delimitrou2014quasar,delgado2015hawk,lo2015heracles}. Previous works such as Paragon~\cite{delimitrou2013paragon} and TetriSched~\cite{tumanov2016tetrisched} have focused on optimizing heterogeneous resource utilization~\cite{rusu2006energy,nishtala2017hipster,narayanan2020heterogeneity,van2012scheduling}, but their resource heterogeneity is pre-determined and sub-optimal, and their target applications are long-running jobs in datacenters, which is different from online inference tasks. Some other previous works have relied on tuning by expertise~\cite{hwang2015cloud,li2010cloudcmp,scheuner2018cloud,shahrad2016availability}, prior profiling~\cite{yadwadkar2017selecting,xu2013bobtail,moghaddam2018energy,zhou2021mocha}, or historical training data from similar applications~\cite{zhang2021sinan,li2022miso,mai2020kungfu,jo2017machine,kim2020autoscale}, and cannot be used to solve the \sol{} problem.

Existing ML inference frameworks ~\cite{lee2020subflow,zhou2018s,yang2019re,gujarati2020serving,crankshaw2017clipper,kannan2019grandslam,wan2020alert,zhang2019mark,ali2020batch,xiang2019pipelined,cui2021enable,romero2021llama,romero2021infaas} are not suitable for exploiting heterogeneous hardware optimally and may require extensive profiling, \sol{} addresses this limitation. For example, S$^3$DNN and DART are heterogeneity-aware deep neural network (DNN) inference frameworks~\cite{zhou2018s,xiang2019pipelined}, but their hardware heterogeneity is pre-determined. INFaaS~\cite{romero2021infaas} selects one particular hardware type from a pool of devices depending on the user application, but unlike \sol{}, it does not explore serving the model using different hardware simultaneously. Media application frameworks such as Llama~\cite{romero2021llama} and Scrooge~\cite{hu2021scrooge} allocate different hardware for different stages of the media application inference, but each query is assigned to the same sequence of hardware types, they do not distribute queries to heterogeneous resources like \sol{} and are not suitable for general purpose applications.

Ribbon~\cite{li2021ribbon} optimizes the serving cost by exploring different heterogeneous configurations, but compared to \sol{}, it still incurs Bayesian Optimization exploration overhead and does not exploit the heterogeneity by intelligently distributing the queries. DeepRecSys~\cite{gupta2020deeprecsys} explores heterogeneity between GPUs and CPUs when serving online queries. However, it does not explore the potential of different CPU/GPU ratios under a cost budget. It uses a hill-climbing algorithm to find an optimal threshold for query distribution, but it incurs tuning overhead as the threshold is different for each heterogeneous configuration. Clockwork~\cite{gujarati2020serving} consolidates design choices in a top-down manner for deterministic inference latencies, but its central controller does not exploit heterogeneous hardware like \sol{}. 
Compared to all previous work, \sol{} delivers a full suite of heterogeneity support for cloud service and considers all key metrics (QoS, throughput, and cost).

\section{Background}
\label{sec:backg}

\noindent\textbf{Machine learning inference service.} When machine learning models are trained into maturity, they will get deployed in production to provide ML inference service. The service users can submit inference requests through provided interfaces (e.g., HTTP request), then get a response. The inference pipeline can have multiple stages (e.g., data pre-processing, model prediction, post-processing), and they are typically packaged into a container image along with the software dependencies. On the cloud, the inference service provider can then allocate a set of compute instances and use a resource manager like Kubernetes to deploy the service. In this work, we focus on discussing the potential of using a heterogeneous resource instance allocation -- how to efficiently distribute the inference queries and find a good heterogeneous configuration quickly.\newline

\noindent\textbf{Inference serving with QoS constraints and cost budget.} The inference service has a QoS target, requiring the tail latency (e.g., $99^{th}$ percentile) of queries to be within a limit for a better user experience. For flexibility reasons and the pay-as-you-go model, businesses rent computing power from the cloud computing provider to meet the QoS target, but they also have a budget constraint. Each compute instance type, rented from the cloud, is associated with a price ($\$/hr$). Given a cost budget, one can only allocate a limited number of instances to serve as many queries as possible -- that is, maximize the query throughput. The query throughput is defined as queries served per second (QPS). Since QoS cannot be violated, we use the \textbf{allowable throughput}, which is the maximum throughput the allocated instances can serve without causing QoS violation. In this work, we use \textit{allowable throughput}, \textit{throughput}, and \textit{QPS} inter-changeably. All of them hold the implicit condition that QoS is satisfied. 

%Allocating cloud instances for maximum throughput under budget is highly desirable for a service provider as more application users can be served concurrently without extra cost.

\section{Motivation}
\label{sec:motiv}

In this section, we first provide experimental evidence to demonstrate that a heterogeneous configuration (a configuration can be a mixture of a few GPU instances, a few instances of CPU type A, and a few instances of type B) \textit{can be} better than a homogeneous configuration \textit{under the same cost budget} while respecting QoS. But, it is not always true -- and any heterogeneous configuration is not superior by simply the virtue of heterogeneity. 

First, we note that given a certain cost budget, one can choose to allocate the most cost-effective instances that can meet the QoS for all queries. We denote such instance type as \textbf{base instance}, and such strategy as \textbf{homogeneous serving} or homogeneous configuration. However, since inference queries have highly diverse batch sizes (or query sizes)~\cite{gupta2020deeprecsys,li2021ribbon,crankshaw2017clipper}, even though a cheaper but higher throughput-per-cost instance type cannot meet the QoS (so it cannot serve standalone as the allowable throughput is 0), it can still meet QoS for some smaller queries (queries with smaller batch sizes) due to the lower latency. Another choice is to replace some base instances with such cheaper instances (denoted as \textbf{auxiliary instances}), we denote this as \textbf{heterogeneous serving} or heterogeneous configuration. Unlike the base instance which comes from the optimal homogeneous instance, multiple types of auxiliary instances can be used for more flexibility and higher potential.\newline

\begin{figure}
    \centering
    \includegraphics[scale=0.5]{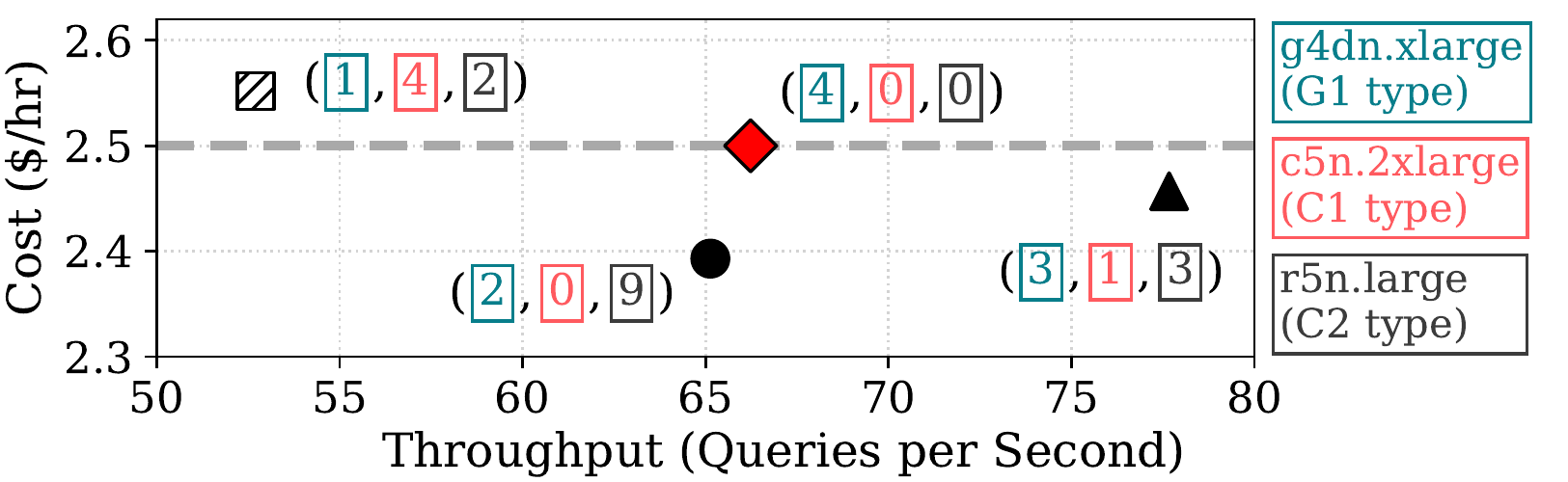}
    \vspace{0.1cm}
    \hrule
    \vspace{-0.4cm}
    \caption{Different heterogeneous configurations versus the best homogeneous one. The number indicates the instance count of each type.}
    \vspace{-0.3cm}
    \label{fig:motiv_1}
\end{figure}

\noindent\textbf{Are heterogeneous configurations always better%(i.e., yield higher throughput)
?} In Fig.~\ref{fig:motiv_1}, we compare the throughput of homogeneous serving against three different heterogeneous configurations on a Meta production model RM2~\cite{gupta2020architectural} under a fixed cost budget (dashed line). All configurations shown here respect the QoS target. We use three AWS EC2 %~\cite{aws_cloud} 
instance types denoted as G1 for base instance, and C1, C2 for auxiliary instances (details in Sec.~\ref{sec:methodology}). The $(4,0,0)$ homogeneous configuration still has some unused budget for 70\% of one G1, so we proportionally scale its throughput and cost up till the budget to give it an advantage.  We observe that heterogeneous outperforms homogeneous as $(3,1,3)$ has 15\% higher throughput than $(4,0,0)$. However, heterogeneity is not always necessarily better (e.g., $(2,0,9)$ and $(1,4,2)$). Especially for $(1,4,2)$, it indicates that simply raising the budget is not an ideal approach to gain throughput. \textit{Therefore, being only heterogeneity aware is not sufficient (like previous work). But, how do we find an optimal configuration like $(3,1,3)$?} \newline

\begin{figure}
    \centering
    \includegraphics[scale=0.47]{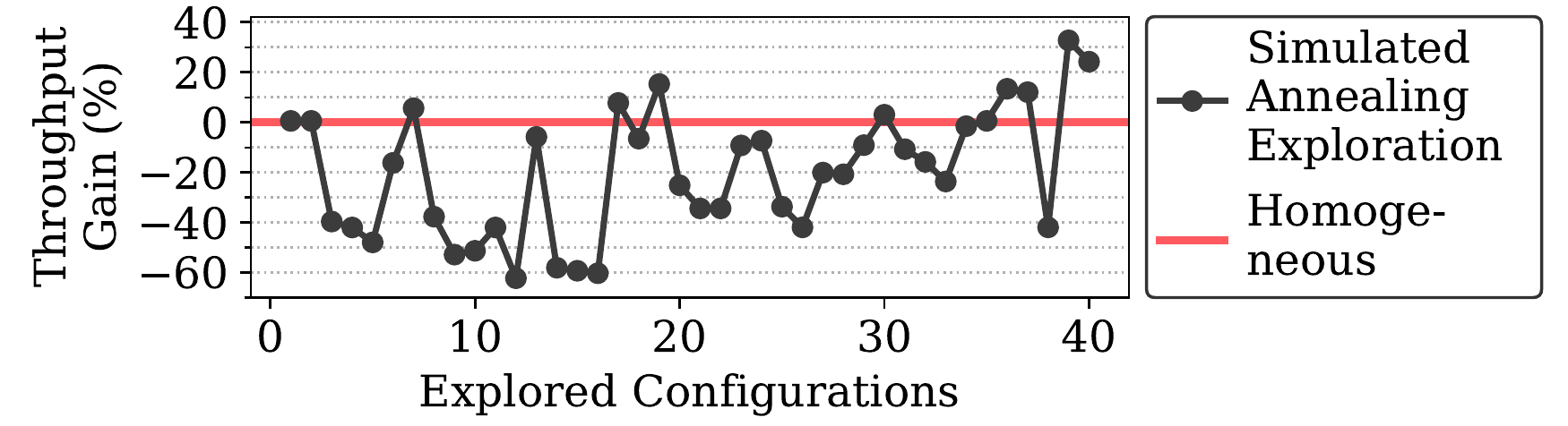}
    \vspace{0.1cm}
    \hrule
    \vspace{-0.4cm}    
    \caption{Throughput improvement over homogeneous when exploring using simulated annealing.}
    \vspace{-0.3cm}
    \label{fig:motiv_2}
\end{figure}

\noindent\textbf{Finding a high-performing heterogeneous configuration is expensive.} This is because the search space of possible heterogeneous configurations is large, especially when there are more instance types, the space becomes high-dimensional and each instance type may have multiple instances. Second, evaluating the throughput of a new configuration is \textit{expensive} and \textit{time-consuming} because it requires service reconfiguration, just allocating new cloud instances would take significant time (tens of seconds). Also, during the online search of configurations, each explored configuration may not yield enough throughput to sustain all the queries - lower throughput than the homogeneous setting. Fig.~\ref{fig:motiv_2} shows the limitation of heterogeneous serving during online exploration using simulated annealing~\cite{van1987simulated}. Although we have pre-filtered out configurations that yield less than 20 QPS, the majority of explored configurations (about 70\%) are still worse than the homogeneous serving marked as the red line. QoS violations will occur frequently if the allowable throughput is below the target level. \textit{High cost of exploring and evaluating has prohibited previous works from finding a better heterogeneous configuration. \sol{} breaks this limitation by providing an approximate method to quickly determine a promising configuration without any online evaluation.}\newline

\begin{figure}
    \centering
    \includegraphics[scale=0.45]{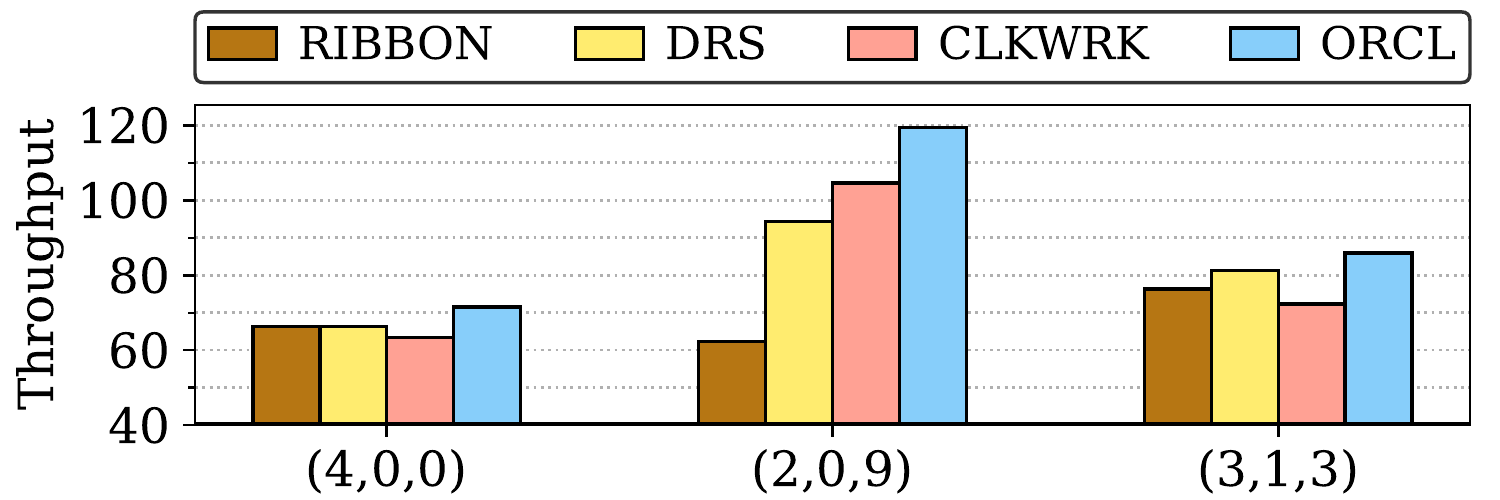}
    \vspace{0.1cm}
    \hrule
    \vspace{-0.4cm}    
    \caption{Heterogeneous configuration performance varies with query distribution mechanism.}
    \vspace{-0.3cm}
    \label{fig:motiv_query}
\end{figure}

\noindent\textbf{Exploiting heterogeneity via intelligent query distribution is the key to higher throughput.} Next, we show that only finding a high-performing heterogeneous is not sufficient. Distributing diverse queries among heterogeneous instances is key to unlocking higher throughput. In previous results (Fig.~\ref{fig:motiv_1} and~\ref{fig:motiv_2}), we used \ribbon{}'s~\cite{li2021ribbon} mechanism to schedule the arrived query on the best instance available. To demonstrate the impact of query distribution strategies, we employ two more complicated query distribution schemes denoted as \drs{}~\cite{gupta2020deeprecsys}, \clk{}~\cite{gujarati2020serving}, and an oracle scheme (ORCL) (details in Sec.~\ref{sec:methodology}) in Fig.~\ref{fig:motiv_query}. We make two observations. First, all state-of-the-art schemes \ribbon{}~\cite{li2021ribbon}, \drs{}~\cite{gupta2020deeprecsys} and \clk{}~\cite{li2021ribbon} are sub-optimal. Second, one is not necessarily better than the other and leaves scope for more improvement. \textit{\sol{} exploits this opportunity and bridges the gap to the oracle scheme by designing a new intelligent query distribution mechanism for heterogeneous serving.}

\section{\sol{} Design}
\label{sec:design}

\begin{figure}
    \centering
    \vspace{0.1cm}
    \includegraphics[scale=0.29]{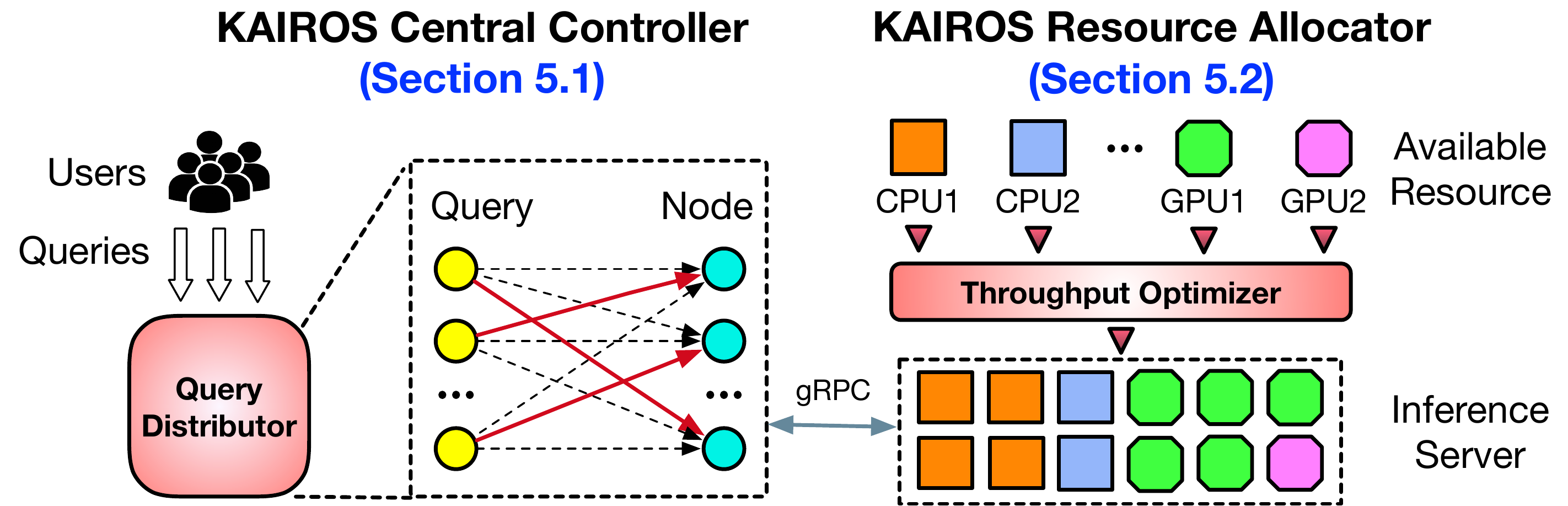}
    %\vspace{-1mm}
    \vspace{0.1cm}
    \hrule
    \vspace{-0.4cm}    
    \vspace{0.0cm}
    \caption{\sol{} Overview.}
    \label{fig:desi_0}
    \vspace{-0.3cm}
\end{figure}

In this section, we provide design details of \sol{} as a complete heterogeneous serving solution, illustrated in Fig.~\ref{fig:desi_0}. The queries submitted by users will be distributed to the processing nodes, which consist of heterogeneous compute instances built by a throughput optimizer. The first design target of \sol{} is to efficiently distribute all the arrived queries of an ML inference service to different instances in a particular heterogeneous configuration (Sec.~\ref{sec:desi_query}). Since finding a promising heterogeneous configuration within a cost budget is also a challenging problem by itself, this is \sol{}'s second design component (Sec.~\ref{sec:desi_throughput}).

\subsection{\sol{} Query Distribution Mechanism}
\label{sec:desi_query}
\noindent\textbf{Overview.} We introduce the query distribution mechanism as the first design component of \sol{}. The key objective is to intelligently distribute queries of different batch sizes to different instances so that the throughput is maximized. In Sec.~\ref{sec:evaluation}, we confirm that \sol{}'s query-distribution mechanism is indeed key to its overall effectiveness and works across different heterogeneous configurations. 

We start with mathematical formulation to maximize the throughput for a given configuration. The key intuition is to distribute the queries in a way that maximizes the available time in all instances in the future. This maximizes the likelihood of serving more queries in the future -- a higher throughput. We show that this problem can be transformed and mapped to a min-cost bipartite matching problem, which \sol{} solves to find an efficient query-distribution plan without knowledge of future query arrivals. \newline

\noindent\textbf{Mathematical formulation of query distribution for throughput maximization.} Our problem objective is: given a number of queries to be served at the current time $t_0$, maximize future availability of instances until a future time instance. This is equivalent to minimizing the total resource usage since unused resources can be used to process future queries, indirectly maximizing throughput at the current time.

\begin{table}[t]
\centering
\vspace{-0.2cm}
\caption{Query distribution optimizer parameters.}
\vspace{-0.3cm}
\scalebox{0.9}{
\begin{tabular}{|c|p{0.45\textwidth}|} 
 \hline
 \multicolumn{1}{|c|}{\textbf{List}} &  \multicolumn{1}{|c|}{\textbf{Description}} \\ [0.5ex] 
 \hline
 $L_{i,j}$ & Time needed to finish serving $Q_i$ on instance $I_j$ from $t_0$. \\ 
 \hline 
 $m$ & Number of queries at time $t_0$. \\ 
 \hline 
 $n$ & Number of instances in the configuration. \\ 
 \hline 
 $C_j$ & Heterogeneity coefficient for instance $I_j$. \\  
 \hline  
 $T_{qos}$ & QoS target latency. \\  
 \hline    
 $W_i$ & Query $Q_i$'s time spent waiting in queue before $t_0$. \\  
 \hline    
 $P_{i,j}$ & Query-to-instance pairing/assignment matrix. \\  
 \hline    
\end{tabular}}
\label{table:desi_distr}
\vspace{-0.3cm}
\end{table}

Suppose at $t_0$, there are $m$ queries in the serving queue, denoted as $Q_1, Q_2, ..., Q_m$, and $n$ compute instances in the heterogeneous configuration, denoted as $I_1, I_2, ..., I_n$. Table~\ref{table:desi_distr} summarizes the parameters used in our mathematical formulation. If distribute query $Q_i$ to instance $I_j$, the query completion time from $t_0$ ($L_{i,j}$) includes the serving latency (varies for different $i,j$) and if there is a query currently being served at $I_j$, the remaining time till $I_j$ can serve $Q_i$. For all queries and instances, this time can be represented by an $m\times n$ matrix $L$. Namely, $L_{i,j}$ represents the $I_j$ instance resource usage (measured by time) if scheduled to serve query $i$.

It is important to note that the equal wall-clock usage time on different instance types in a heterogeneous configuration do not hold the same value. That is, one second of GPU is not equivalent to one second of CPU. To account for this, \sol{} employs a heterogeneity coefficient $C_j$ for each instance type $j$. 
\begin{definition}
We define $C_j \in (0,1]$ as the heterogeneity coefficient for instance $j$. It is used to represent the relative importance of instance $j$ compared to other instances in a heterogeneous system. It is calculated as the ratio of the largest query latency between $I_j$ and the base instance type.
\end{definition}
\revision{The heterogeneity coefficient helps \sol{} weight resources differently, which aligns with previous task scheduling algorithms for heterogeneous processors~\cite{orr2021optimal,topcuoglu2002performance}.} To determine $C_j$, we first set the coefficient of the base instance type (e.g., lowest latency instance) to 1 as a normalization point, then calculate $C_j$ as the latency ratio. We find that using the largest query the system can serve to measure the latency ratio works well. For example, if the largest query has latency 100ms on instance $I_1$, 200ms on $I_2$ and 500ms on $I_3$, then $C_1 = 1$, $C_2 = 0.5$, $C_3 = 0.2$. In our system, we limit the maximum batch size of a query to 1000 because of QoS constraints. Intuitively, larger queries are more compute demanding and more prone to violate QoS, thus, they are more suitable for evaluating the relative importance of instances in a heterogeneous system.
% For example, if a query latency on $I_j$ is 40ms and 20ms on the base instance, then, $C_j$ is equal to $0.5$.  
With the introduction of heterogeneity coefficient, the \textbf{revised usage time} for instance $j$ can be expressed as $C_jL_{i,j}$. 

This usage can be calculated for every query/instance pair. Since the time is relative to the base instance, we can directly sum the usage up across all instances, and the sum is the aggregated resource usage. To minimize this, we need to carefully select which $Q_i$ to be served on which $I_j$. We define these optimization variables as an $m\times n$ binary matrix $P$:
{
\begin{align}
P_{i,j} =     
    \begin{cases}
        1 & \text{if query $Q_i$ is served by instance $I_j$,}\\
        0 & \text{otherwise}. \\
    \end{cases}
\end{align}}%
Then, we express the minimization objective function as:
{
\begin{align}
\label{eq:desi_gx}
f(P) = \sum_{i=1}^{m} \sum_{j=1}^{n} C_jL_{i,j}P_{i,j}
\end{align}}%
\textit{Before optimizing this objective function, we first make an observation that this formulation maps closely itself to the linear-sum assignment problem, or in graph theory, a min-cost bipartite matching problem~\cite{karp1990optimal}. Therefore, we leverage the theory of bipartite matching to formulate and solve this as a bipartite matching optimization.}

A bipartite graph has two disjoint and independent sets of vertices $U$ and $V$. In our case, $U$ contains the queries as vertices, $V$ contains the instances as vertices. An edge is available for all query-instance pairs, and a cost is associated with each edge. A typical min-cost bipartite matching would have the same number of elements in $U$ and $V$, the elements are one-to-one matched with the total cost minimized. However, in our situation, there is no guarantee about the number of queries. If there are fewer queries than instances, the matching is \revision{valid} when all queries are matched to a unique instance, and when there are fewer instances than queries, \revision{the matching is valid when all instances are matched to a unique query}. The cost of each edge between $Q_i$ and $I_j$ corresponds to $C_jL_{i,j}$ in Eq.~\ref{eq:desi_gx}.

Before solving this bipartite matching problem to maximize the throughput, we note that processed queries can only count towards throughput when served under QoS, otherwise \sol{}'s idea of heterogeneous serving becomes meaningless: one can simply find the instance type with the highest throughput-to-cost ratio and do homogeneous serving. To be QoS-aware, we add an inequality constraint:
{
\begin{align}
\label{eq:desi_qos}
(L_{i,j} + W_i)P_{i,j} \leq T_{qos}
\end{align}}%

This constraint states that if serving $Q_i$ with $I_j$, the sum of query completion time on $I_j$ and queue wait time before $t_0$ should be less than the QoS target. We need to consider the query wait time $W_i$ because not all queries are guaranteed to be scheduled to an instance (e.g., more queries than instances), they may need to wait in a queue until more resources become available and restart another round of query distribution. Considering $W_i$ in Eq.~\ref{eq:desi_qos} avoids starvation of unscheduled queries when new queries continuously arrive. 

\begin{figure}
    \centering
    \includegraphics[scale=0.21]{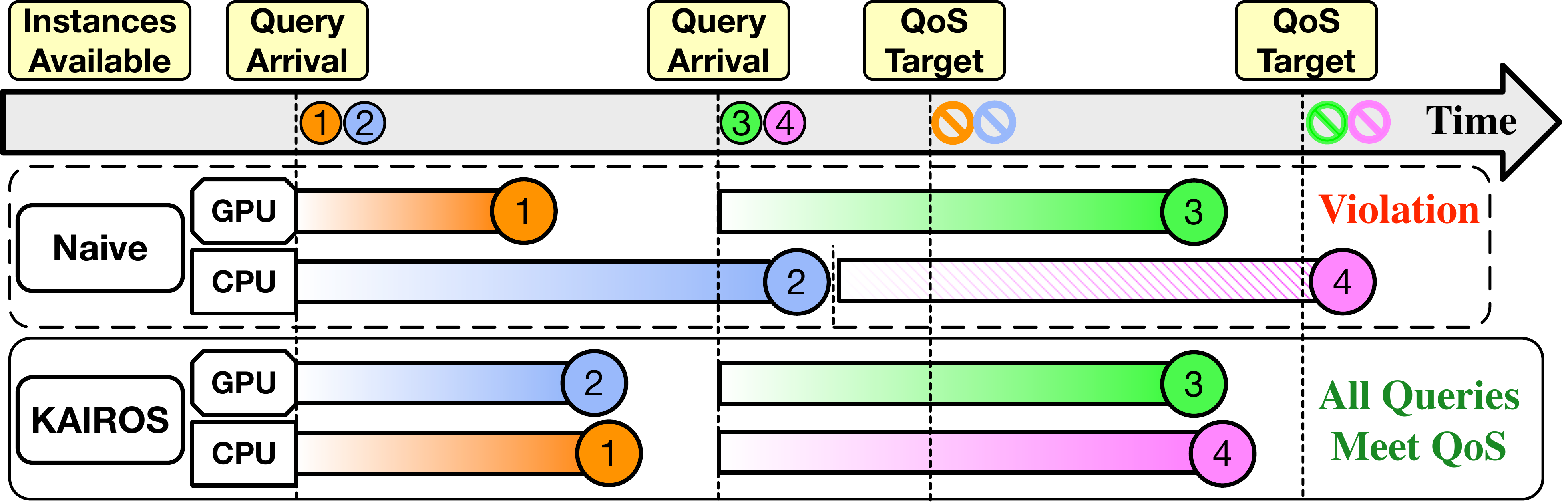}
    % \vspace{-1mm}
    \vspace{0.1cm}
    \hrule
    \vspace{-0.4cm}    
    \vspace{0.0cm}
    \caption{Higher-speedup queries should be prioritized on more powerful instances to create more slack time. Queries 1 to 4 (arrived in order) are represented in different colors.}
    \label{fig:desi_1}
    \vspace{-0.3cm}
\end{figure}
In summary, \sol{} formulates an optimization problem and designs its objective function and constraints for throughput maximization under the QoS target. The key design principle is respecting the fact that different queries have different speedups from one instance type to another (e.g., queries with larger batch sizes have higher speedups from CPU to GPU). By prioritizing higher speedup queries on more powerful instances, \sol{} minimizes resource usage and prepares maximized slack time for future queries. This is reflected in Eq.~\ref{eq:desi_gx}, where the $L$ matrix implicitly contains this information. Fig.~\ref{fig:desi_1} visualizes this effect with a 2-instance example. By efficiently distributing current queries without future information, \sol{} leaves more slack time for the future, thus it can process all 4 queries while a naive scheme (e.g., FCFS) can only process 3 queries (shaded query does not count towards throughput due to QoS violation). 
The superior distribution scheme gives \sol{} 33\% higher throughput (4 queries vs. 3 queries processed in time) than the naive scheme, despite the same hardware.

Putting everything together, the query-distribution problem can be formulated as follows:
{
\begin{align}
    \label{eq:desi_obj}
    &\min_{P}
    \begin{aligned}[t]
      & \sum_{i=1}^{m} \sum_{j=1}^{n} C_j(L_{i,j})P_{i,j}
    \end{aligned} \\
    & \label{eq:desi_c0}\text{s.t. } \;\forall i, j,\;\; (L_{i,j} + W_i)P_{i,j} \leq T_{qos}, \\
    & \label{eq:desi_c1} \;\;\;\;\;\; \forall i, j, \;\; \sum_{i=1}^{m} P_{i,j} \leq 1, \sum_{j=1}^{n} P_{i,j} \leq 1, \\
    & \label{eq:desi_c2} \;\;\;\;\;\; \sum_{i=1}^{m} \sum_{j=1}^{n} P_{i,j} \geq min\{m, n\}
\end{align}}%
where $i \in \{1, 2, ..., m\}$ and $j \in \{1, 2, ..., n\}$. Eq.~\ref{eq:desi_c1} indicates one-one mapping, and Eq.~\ref{eq:desi_c2} guarantees when there are more instances than queries, every query gets mapped to an instance; when there are more queries than instances, every instance receives a query.

We note that \sol{}'s formulated problem is not a strict min-cost bipartite matching problem as discussed in traditional bipartite matching literature because of the QoS constraint in Eq.~\ref{eq:desi_c0}. Therefore, to guarantee feasibility, \sol{} integrates this constraint into the objective function by modifying the $L$ matrix with a condition. If serving $Q_i$ on $I_j$ does not violate the QoS, $L_{i,j}$ is unchanged. If it violates QoS, then $L_{i,j}$ is penalized by a large quantity (e.g., $10\times$ of the QoS target). Consequently, \sol{} achieves min-cost solutions that avoid QoS-violating $Q_i$-$I_j$ matching. With this constraint integration, the new $L$ matrix becomes:
{
\begin{align}
L_{i,j} = 
    \begin{cases}
        L_{i,j} & \text{if Eq.~\ref{eq:desi_c0} is true,} \\
        10\cdot T_{qos} & \text{otherwise.}
    \end{cases}
\end{align}}%

Then, the Eq.~\ref{eq:desi_c0} constraint is removed, and the problem with updated parameter $L$ becomes a strict min-cost bipartite matching problem. \sol{} solves this problem using the Jonker-Volgenant algorithm~\cite{jonker1987shortest} which is a variant of the widely used Hungarian algorithm~\cite{kuhn1955hungarian}, but more efficient in practice~\cite{crouse2016implementing}.\newline

\noindent\textbf{Remarks on assumptions and overhead.} We note that \sol{} requires constructing the parameter matrix $L$, which requires predicting the query latency of certain batch sizes on different instance types. Fortunately, ML inference is a fully deterministic process without conditional branching, thus the latency is highly predictable~\cite{gujarati2020serving}. Because the query includes a batch of requests, \sol{} makes sure an instance serves one query at a time without any resource contention. Thus, the end-to-end latency has a very low variance ($<0.5\%$ of mean). Previous work has observed that inference latency can be accurately predicted with simple features such as request batch size~\cite{chen2022retail}. We have observed similar trends in our experiments as inference latency is highly correlated with query batch size: the Pearson correlation coefficient~\cite{benesty2009pearson} between latency and batch size is greater than 0.99 for all models and instance types in Sec.~\ref{sec:methodology}. 
As a result, \sol{} can model this completely online with a handful of queries without requiring any prior knowledge or instrumentation. \sol{} starts with a linear model but does not rely on the model accuracy because it will quickly transition into a lookup table after processing more queries. All our evaluation results include this overhead from learning the query latencies online.
In practice, as a noise safeguard, we replace $T_{qos}$ with $\xi T_{qos}$ in Eq.~\ref{eq:desi_c0}, and setting $\xi$ to be 0.98 such that the completion time predicted to be within 2\% range of the QoS target is considered a violation.

\subsection{\sol{} Throughput Estimation}
\label{sec:desi_throughput}

Next, we discuss how \sol{} quickly reaches a good configuration from the vast search space with a cost budget constraint as the second part of its design component. Calculating the cost is easy but evaluating the throughput of a configuration is expensive and causes delays in finding a good configuration (Sec.~\ref{sec:motiv}), prohibiting the system from promptly responding to load changes. \sol{} takes a different approach to approximate the actual throughput using an upper bound. \revision{Classical approximation algorithms for unrelated machines~\cite{lenstra1990approximation,azar2005convex} are not applicable to our application scenario of serving online inference queries using \sol{}'s query distribution mechanism.} We have also explored other options such as queuing theory~\cite{lehoczky1996real,cooper1981queueing} to analytically calculate the actual throughput. However, due to the dynamic service time (varying batch size), the heterogeneity in hardware, and unconventional queue discipline (Sec.~\ref{sec:desi_query}), we cannot fit the problem into a classical $M/M/c$ queue framework. Therefore, we take \sol{}'s approximation approach to avoid expensive evaluations. 

Designing an application-specific approximation strategy is challenging~\cite{mittal2016survey}. This is also true for \sol{}'s throughput approximation due to QoS restrictions and the complex interactions between queries and heterogeneous instances. \sol{} tackles this challenge with a method to calculate a throughput upper bound for a given configuration. With this method, \sol{} first finds promising candidates with high upper bounds from search space under cost budget without any evaluation, then performs aggregation to output a final configuration. To explain, we first demonstrate the intuition and calculation for the throughput upper bound given a simple heterogeneous configuration -- that is, one instance of base type (e.g., GPU) and one instance of auxiliary type (e.g., CPU). Then, we show how to extend this method to cases where each instance type can have multiple instances. Eventually, we extend it to cases where one can have multiple different types of auxiliary instances. \begin{definition}
For an inference service, given a particular allocation of hardware resources, the throughput QPS varies with the query distribution algorithm $\lambda$. The allowable throughput can be represented as a function $Q(\lambda)$. We define the throughput upper bound $QPS_{max}$ as a number that satisfies $\forall \lambda$, $Q(\lambda) < QPS_{max}$.
\end{definition}

Essentially, the upper bound of a particular hardware allocation is a throughput that cannot be exceeded no matter how the system distributes the query. A throughput number may be the upper bound for a hardware allocation, but if the allocation changes, it may not be the upper bound anymore.

We estimate the throughput upper bound as follows: \sol{} makes a simple observation that upper bound estimation is akin to estimating the maximum possible throughput in an unrealistic scenario where all queries are available to us at the beginning, and we can control when each query should arrive -- then there is no need to worry about latency interactions with queuing. Recall that \sol{}'s query distribution mechanism efficiently accounts for the practical case when we have no control over when queries will arrive: queries may wait in the queue, and instances may be idle waiting for queries. Compared to the practical case, our upper bound calculation ensures queries do not miss QoS by waiting in the queue and instances do not waste idle cycles that could have served more queries.

Using the one-base-one-auxiliary simplification example, the intuition is to determine which instance type is the bottleneck given a mixture of queries \revision{of various batch sizes}, and then, the bottleneck \revision{instance} type dictates the maximum possible throughput we can achieve. Formally, let $Q_{b}$ and $Q_{a}$ denote the \textit{standalone} throughput (QPS) achieved by the base and auxiliary instance type, respectively. Note that the \revision{standalone} auxiliary instance cannot satisfy QoS for all queries. Therefore, $Q_{a}$ refers to the throughput achieved when serving only queries that do not violate QoS (i.e., queries smaller than size, say $s$). Queries larger than size $s$ will then have to be served by the base instance. However, the base instance throughput when serving larger-than-$s$-size queries will be lower than $Q_{b}$ because larger batches require longer processing time -- we use $s+$ to represent queries larger than size $s$ (that cannot be consumed by auxiliary instance) and denote their allowable throughput running on base instances by $Q^{s+}_{b}$. 

\begin{figure}
    \centering
    \includegraphics[scale=0.37]{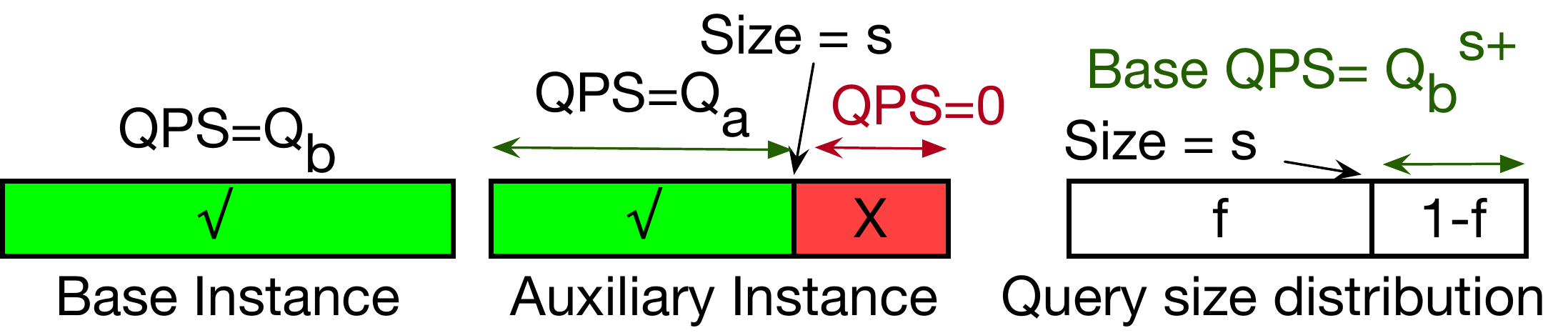}
    \vspace{0.1cm}
    \hrule
    \vspace{-0.4cm}    
    \caption{Upper bound calculation parameters. The auxiliary instance cannot serve larger queries due to QoS.}
    \label{fig:desi_2}
    \vspace{-0.3cm}
\end{figure}

We can then partition the query mix into two fractions: $f$ for queries smaller than batch size $s$, and $1-f$ for queries larger than $s$. If the auxiliary instance is fully occupied with smaller size queries, it implies that $\frac{1-f}{f}\times Q_{a}$ of \revision{the throughput needs to be executed on the base instance type, represented by the ratio between $s+$ queries and the $Q_a$ queries times the $Q_a$ throughput}. Sending larger queries to base and small queries to auxiliary instances also aligns well with the design principle of ~Sec.~\ref{sec:desi_query}. However, recall that all the queries \revision{offloaded to base instance} are of sizes larger than $s$, and hence, the base instance can serve them only at the rate of $Q^{s+}_{b}$. Fig.~\ref{fig:desi_2} shows a visual representation of the mathematical formulation above.

Note that there are only two potential outcomes that dictate the maximum throughput we can achieve: (1) the base instance is the bottleneck, and (2) the auxiliary instance is the bottleneck.

If $Q^{s+}_{b} \leq \frac{1-f}{f}\times Q_{a}$, that means \textit{the base instance is the bottleneck}: \revision{the queries offloaded from auxiliary instance cannot be fully consumed by the base instance, thus the auxiliary instance cannot serve queries at the rate $Q_{a}$.} If the maximum throughput is denoted as $QPS_{max}$, then, a $1-f$ fraction of \revision{all the} queries (or queries larger than size $s$) will be served by the base instance at the rate of $Q^{s+}_{b}$. Hence, the maximum throughput can be estimated as:
{
\begin{align}
\label{eq:desi_base}
QPS_{max} = \frac{Q^{s+}_{b}}{1-f}
\end{align}}%
If $Q^{s+}_{b} > \frac{1-f}{f}\times Q_{a}$, that means \textit{the auxiliary instance is the bottleneck and the base instance has some slack left to serve more queries}. If the maximum throughput is denoted as $QPS_{max}$, then, $f$ fraction of these queries (queries smaller than size $s$) will be served by the auxiliary instance at the rate of $Q_{a}$. So, the $QPS_{max}$ is equal to $\frac{Q_{a}}{f}$. However, recall that the base instance still has some slack available to serve more queries. Intuitively, the slack ratio at the base instance is equal to the difference between $Q^{s+}_{b}$ and $\frac{1-f}{f}\times Q_{a}$, divided by its total capability $Q^{s+}_{b}$. Multiplying this slack ratio by the base throughput, we get the base slack throughput (extra queries the base can serve when the auxiliary is the bottleneck) as:
{
\begin{align}
\text{Base-slack-throughput} = (\frac{Q^{s+}_{b} - \frac{1-f}{f}\times Q_{a}}{Q^{s+}_{b}})\times Q_{b}
\end{align}}%

Therefore, the maximum possible throughput is: 
{
\begin{align}
\label{eq:desi_aux}
QPS_{max} = \frac{Q_{a}}{f} + (\frac{Q^{s+}_{b} - \frac{1-f}{f}\times Q_{a}}{Q^{s+}_{b}})\times Q_{b}
\end{align}}%

\begin{figure}
    \centering
    \includegraphics[scale=0.286]{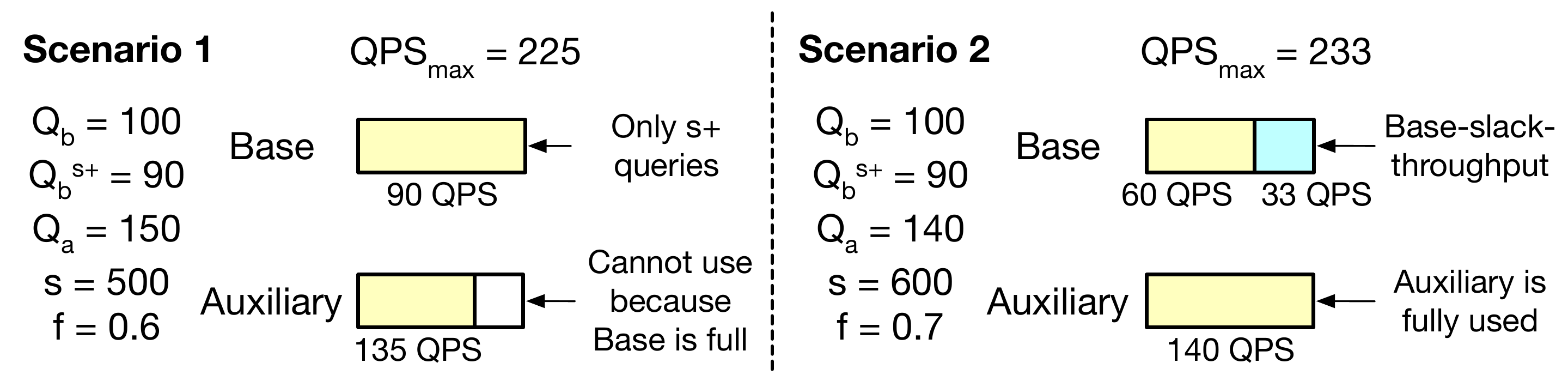}
    \vspace{0.1cm}
    \hrule
    \vspace{-0.4cm}    
    \caption{\revision{Example of how \sol{}'s upper bound calculation works. Scenario 1 represents when the base instance is the bottleneck and Scenario 2 represents when the auxiliary instance is the bottleneck.}}
    \label{fig:desi_new}
    \vspace{-0.3cm}
\end{figure}

\revision{Fig.~\ref{fig:desi_new} uses two example scenarios 1 and 2 to demonstrate the upper bound calculations. In Scenario 1, the auxiliary instance can only process queries smaller than size 500, and the base has to spend all its capacity processing the $Q^{s+}_{b}$ queries, \sol{} uses Eq.~\ref{eq:desi_base} to calculate $QPS_{max}$. In Scenario 2, the base instance still has some slack capacity left after processing the $Q^{s+}_{b}$ queries, and \sol{} uses Eq.~\ref{eq:desi_aux} to calculate $QPS_{max}$.} 

\sol{}'s approach of query mixture partition \revision{by batch size $s$} would be over-optimistic if the partitions have strong temporal locality (e.g., if all large queries arrive together before small queries, then auxiliary instance cannot contribute). However, no such cyclic behavior has been observed, and due to the law of large numbers, this upper bound is reasonable over the long term for a large number of query mixes. 
We can next extend this method to the case when each instance type has multiple nodes. If the base instance has $u$ nodes and the auxiliary instance has $v$ nodes, then, Eq.~\ref{eq:desi_base} and~\ref{eq:desi_aux} can be written as:
{
\begin{align}
QPS_{max} = \frac{uQ^{s+}_{b}}{1-f}
\end{align}
\begin{align}
QPS_{max} = \frac{vQ_{a}}{f} + (\frac{uQ^{s+}_{b} - \frac{1-f}{f}\times vQ_{a}}{uQ^{s+}_{b}})\times uQ_{b}
\end{align}
}%
The next step is to extend this upper bound estimation for multiple types of auxiliary instances. Multiple types of auxiliary instances are more challenging since each new auxiliary instance has its own QoS-respecting region and throughput in that region. Fortunately, \sol{} is not concerned with modeling the accurate throughput, instead, it only cares about the upper bound estimation of the throughput. Therefore, it makes a relatively simple approximation that additional auxiliary instance types have the same QoS-respecting region as the type with maximum $s$ size and $f$ fraction. This essentially makes the upper bound estimation more optimistic since some weaker auxiliary instances are assumed to meet QoS for batch sizes larger than their limit. As our evaluation confirms (Sec.~\ref{sec:eval_why}), even though this approximation results in a higher upper bound, configurations still follow similar order as the actual throughput (a higher upper bound is likely to indicate higher throughput). With this simplification, the \revision{$n$-auxiliary-instance-type} general case upper bound can be written. First, we define an intermediate variable:
{
\begin{align}
\label{eq:constr}
C = \frac{\sum_{i=1}^{n}v^iQ^i_a(1-f')}{f'}
\end{align}}
where $f'=max(f^1,f^2,...,f^n)$. This corresponds to $\frac{1-f}{f}\times Q_{a}$ that is used to compare with $Q^{s+}_b$. The superscript $i$ indicates auxiliary instance type $i$. $Q^i_a$ is the instance type $i$ throughput for queries with batch size smaller than the maximum $s$ of all types. We have the final upper bound formula as:
{
\begin{align}
    QPS_{max} = \begin{cases}
        \frac{uQ^{s+}_{b}}{1-f'} & \text{if } uQ^{s+}_b \leq C,\\
        \frac{\sum_{i=1}^{n}v^iQ^i_a}{f'} + (\frac{uQ^{s+}_{b} - C}{uQ^{s+}_{b}})uQ_{b} & \text{otherwise.} \\
        \end{cases}
        \label{eq:ub}
\end{align}}%

Finally, now that we have the formula, we can calculate the upper bound for all configurations within the cost budget. We describe the last step that \sol{} performs to reach the final configuration without evaluations. To this end, \sol{} can trivially pick the configuration with the highest upper bound from its approximation method. However, \sol{} recognizes that a higher upper bound does not necessarily mean a higher throughput. To address this, \sol{} applies a similarity-based method to pick a configuration from the highest upper bound configurations. \sol{} first checks if the top-3 upper bound configurations have the same base instance number. If true, \sol{} picks the highest upper bound configuration. Otherwise, for each configuration with a top-10 upper bound, \sol{} calculates its squared Euclidean distance to the other 9 configurations, sums them up, and picks the one with the least distance sum. Such metric is commonly used in clustering analysis~\cite{nainggolan2019improved}, it is equivalent to considering all configurations to form a cluster, then setting the cluster centroid as the configuration that has the least sum-of-squared error (SSE). The intuition is that there should be a region for the high throughput configurations, and the distance-based method lands \sol{} in such a region. We find Euclidean distance to be a reasonable similarity metric, other metrics such as cosine similarity do not reflect the locality of the promising region. As our evaluation confirms (Sec.~\ref{sec:evaluation}), \sol{} is able to find a good configuration for all workloads, and the process does not evaluate any configuration online. \newline

\noindent\textbf{Remarks on assumptions and overhead.} \sol{}'s upper bound based throughput estimation has a one-time warmup phase consisting of two major steps. Firstly, it needs to compute the upper bound for all combinations of instance numbers of each type under cost budget. Fortunately, this calculation is quick using the formula in Eq.~\ref{eq:ub}: for an order of 1000-configuration search space, all upper bounds can be calculated and ranked within 2 seconds, negligible compared to even one evaluation (tens of seconds for instance allocation). Secondly, \sol{}'s estimation implicitly assumes that it can obtain information on the batch size distribution (fraction $f$ of batch sizes smaller than $s$). This is done via query monitoring to keep track of a number of most recent queries (e.g., 10000 queries), and does not require extra profiling. In addition, to provide robustness, we demonstrate that \sol{} adapts when the batch size distribution changes and continues to be effective (Sec.~\ref{sec:eval_transient}). \newline

\noindent\textbf{Upper-bound-assisted search algorithm.} We also develop \solplus{}, a variation of \sol{} that uses a minimum number of online evaluations to quickly find the optimal (Algorithm~\ref{algo:greedy}). The search process is guided by the estimated upper bounds and is shown to be outperforming any other traditional search space exploration (Sec.~\ref{sec:evaluation}). 
The intuition to greedily start from high upper bound configurations as these configurations have better potential than others, and after evaluating a number of such instances, the current best throughput will likely be high enough to filter out a large number of configurations whose upper bounds are lower. Another pruning mechanism in the algorithm is sub-configuration pruning. If configuration $\bm{x_1}$ can add more instances to become $\bm{x_2}$, we define $\bm{x_1}$ to be a sub-configuration of $\bm{x_2}$. Every time a configuration is evaluated, all of its sub-configurations get pruned away from the search space since these sub-configurations will not have higher throughput than the evaluated one. We note that upper bounds that are tight to the actual throughput are especially beneficial to \solplus{} since more configurations can be pruned away. 

\begin{algorithm}[t]
\SetAlgoLined
$UBs \leftarrow $ Sort all $QPS_{max}$ high to low \\
$curr\_best = 0$ \tcp{Highest throughput so far}
$best\_config = None$ \\
$configs \leftarrow $ list of all configs within cost budget \\
$\bm{x} \leftarrow $ variable representing one configuration \\
\ForEach{$UB(\bm{x})$ \text{\textbf{in}} $UBs$} 
{
    \If{$\bm{x} \in configs$}{
    $eval = f(\bm{x})$ \tcp{Actual QPS evaluation.}
    \If{$eval > curr\_best$}{
        $curr\_best = eval$ \\
        $best\_config = \bm{x}$ \\
        Filter all $\bm{c}$ out of $configs$ that satisfies $UB(\bm{c}) \leq curr\_best$ \\
    }
    Prune away all sub-configs. of $\bm{x}$ from $configs$ \\}
}
\Return $curr\_best$, $best\_config$
\caption{\solplus{}'s pruning-based algorithm for quickly finding optimal configuration.}
\label{algo:greedy}
\end{algorithm}

\section{Implementation}

\sol{} and \solplus{} are implemented as a cloud inference server, similar to frameworks such as NVIDIA Triton~\cite{triton}. 
Every allocated compute instance hosts a copy of the model, only one query consisting of the batched requests is served by one model copy at a time. If the model is hosted on a CPU instance, all the CPU cores will be used. The query distribution mechanism (Sec.~\ref{sec:desi_query}) resides in a central controller, which performs the optimization to decide the query-instance mapping. It acts as a client and sends the optimized inference requests to individual instances (as servers) through the gRPC protocol~\cite{grpc}. Compared to a traditional load balancer, the central controller is aware of the server heterogeneity and tries to avoid QoS violations. When queries arrive, the controller estimates the query latency and uses its estimated server remaining time to construct the $\bm{L}$ matrix for Eq.~\ref{eq:desi_obj} and~\ref{eq:desi_c0}.

Notice that the central controller needs to solve the optimization problem in real time because the solving time is added to the inference latency. We implement this using the \texttt{scipy.optimize} package to solve the bipartite matching problem in polynomial time. We experimentally confirmed that the sum of network delay and algorithm runtime for a large 20-query-20-instance matching is within 0.05ms, and even for hundreds of queries arriving concurrently the overhead is well within 1ms, negligible compared to QoS which is typically tens to hundreds of milliseconds. Thereby, \sol{} ensures that its controller does not become the bottleneck or add significant latency. Furthermore, according to the theory from POP~\cite{narayanan2021solving}, inference service frameworks like \sol{} can scale to extremely large systems by dividing the system into multiple sub-systems and running a \sol{} instance on each sub-system.

\section{Methodology}
\label{sec:methodology}

\noindent\textbf{Models and QoS constraints.} We use industry-scale machine learning service models to drive the evaluation of \sol{}' effectiveness. Such ML models are widely used in online services and have several advantages: (1) wide interest from systems research~\cite{li2021ribbon,kalamkar2020optimizing,xie2020kraken} (2) large customer demand and wide deployment in industry~\cite{naumov2019deep,argyriou2020microsoft} (3) representative public trace and artifacts for reproducibility and comparison~\cite{gupta2020deeprecsys,gupta2020architectural,hsia2020cross}. Table~\ref{table:metho_model} lists the models and QoS as 99$^{th}$ tail latency target based on their service specifications.

\begin{table}[t]
\centering
\vspace{-0.2cm}
\caption{Models and QoS targets.}
\vspace{-0.3cm}
\scalebox{0.82}{
\begin{tabular}{p{0.09\textwidth}p{0.18\textwidth}p{0.17\textwidth}p{0.05\textwidth}} 
\toprule
\textbf{Model} & \textbf{Description} & \textbf{Application} & \textbf{QoS}\\ 
\midrule
\midrule
NCF~\cite{he2017neural} & Collaborative Filtering & Movie recommendation & 5 ms \\ 
\hline  
RM2 \cite{gupta2020architectural} & Meta's recommendation model class 2 & High-accuracy social media posts ranking & 350 ms\\ 
\hline  
WND \cite{cheng2016wide} & Google Wide and Deep recommender system & Google App Store & 25 ms \\  
\hline   
MT-WND \cite{zhao2019recommending} & Multi-Task Wide and Deep, predicts multiple metrics in parallel & YouTube video recommendation & 25 ms \\  
\hline  
DIEN \cite{zhou2019deep} & Alibaba Deep Interest Evolution Network & E-commerce & 35 ms \\  
\bottomrule
\end{tabular}}
\label{table:metho_model}
\vspace{-0.3cm}
\end{table}

These models are chosen because they represent a wide range of ML-based applications, the internal architectures are also highly diverse across models~\cite{gupta2020architectural,hsia2020cross}. For example, NCF is a light-weighted model with limited embedding tables, but the RM2 model is dominated by large embedding tables, while the MT-WND model has large parallel DNN predictors for abstract features. The QoS constraints for these models cover a wide range and are selected strictly based on real applications~\cite{cheng2016wide,gupta2020deeprecsys,zhou2019deep}. For example, Alibaba's e-commerce service requires tens of milliseconds of response time while Meta's social media platform requires hundreds of milliseconds.

An important feature of model queries is that they arrive in different batch sizes (Sec.~\ref{sec:motiv}). Our evaluation is driven by the production trace of real query batches from Meta~\cite{gupta2020deeprecsys}. The query inter-arrival is generated from a Poisson process generating 100s of queries per second, which has been commonly used in various online inference serving studies~\cite{gupta2020architectural,gan2019open,reddi2020mlperf,kasture2016tailbench}. To evaluate \sol{}'s response to load change and sensitivity, we also use Gaussian distributed batch sizes because Gaussian distribution is another commonly used distribution for online services~\cite{li2016work}.\newline

\noindent\textbf{Computing hardware types and cost.} We use different hardware types provided by Amazon Elastic Compute Cloud (EC2) for our evaluation. The compute instances are categorized into four classes: general purpose, compute optimized, memory optimized, and accelerated, and represent different cost points. We select an instance type of each class to form the heterogeneous pool, Table~\ref{table:metho_instance} summarizes the compute instance types. These instance types are selected to collectively cover a wide spectrum of performance and cost points, as they come from different representative compute-memory-accelerator classes.

\begin{table}[t]
\centering
% \vspace{-0.1cm}
\caption{Different instance types used in heterogeneous pool.}
\vspace{-0.3cm}
\scalebox{1}{
\begin{tabular}{ccc}
\toprule
\textbf{Instance Type} & \textbf{Instance Class} & \textbf{Price (\$/hr)}\\ 
\midrule
\midrule
{g4dn.xlarge} & GPU Accelerated Computing & 0.526 \\ 
\hline  
{c5n.2xlarge} & Compute Optimized CPU & 0.432 \\ 
\hline  
{r5n.large} & Memory Optimized CPU & 0.149 \\ 
\hline  
{t3.xlarge} & General Purpose CPU & 0.1664 \\ 
\bottomrule
\end{tabular}}
\label{table:metho_instance}
\vspace{-0.3cm}
\end{table}

These instance types create a 4-dimensional search space, each heterogeneous pool corresponds to a certain number of instances of each type. We select the instance size (indicated by \texttt{.xlarge} characters) so that all types have 16GB of memory allocation to host the model. We use the \texttt{g4dn.xlarge} GPU instance type as the base instance type as only this instance type can meet QoS for all batch sizes. The other instance types (e.g., c5n.2xlarge) are considered auxiliary instance types. \texttt{g4dn.xlarge} instance type uses one NVIDIA T4 GPU. Compared to all other GPU-accelerated instance types in EC2, \texttt{g4dn} has the best inference performance -- similar to \texttt{p3.2xlarge} instance type (NVIDIA V100 GPU). But, \texttt{p3} has nearly $6\times$ higher cost than \texttt{g4dn}, therefore, we use \texttt{g4dn} as the most cost-effective base instance type to be used in a homogeneous pool. The cost budget is set to  $2.5 \$/hr$ by default for \sol{} evaluation - this budget is chosen purely to ensure the cost incurred during the evaluation period remains within a reasonable limit while demonstrating the value of the idea. As our evaluation confirms, \sol{} is not sensitive to the cost budget (e.g., $10 \$/hr$ budget) and is not tuned to work specifically only for certain cost budget caps. \sol{} respects the budget cap as a constraint -- consistent with our design goal. \newline

\noindent\textbf{Metrics.} Our main evaluation metric remains the throughput (QPS) under QoS (Sec.~\ref{sec:motiv}). To find this allowable throughput, we gradually increase the arrival rate of queries, until the QoS is violated. In addition, since evaluating a configuration in the search space is expensive, we also compare the number of iterations required to find the optimal configuration for a particular scheme. This part is only applicable to competing techniques (non-\sol{} solutions) since \sol{} does not require online exploration and evaluation.\newline

\noindent\textbf{Competing query distribution techniques.}  We evaluate \sol{} against three recent ML inference serving schemes. \vspace{1mm}

\noindent\textbf{\underline{\ribbon{}}.} This scheme focuses only on hardware allocation, therefore it has a simple first-come-first-serve (FCFS) query distribution policy which prefers instances of the base type when multiple instances are available. However, it uses Bayesian optimization to allocate a near-optimal set of heterogeneous hardware on the cloud, which is similar to the problem \sol{} tries to solve in Sec.~\ref{sec:desi_throughput}. \vspace{1mm}

\noindent\textbf{\underline{\drs{}}.} This scheme represents the scheme used in a related work -- DeepRecSys system~\cite{gupta2020deeprecsys}. It uses a static batch size threshold to decide whether to serve a query on the base instance (if large than a threshold) or on the auxiliary instance (otherwise). To determine the threshold, a hill-climbing sweep is used by DeepRecSys system~\cite{gupta2020deeprecsys} to find the threshold that yields the highest throughput.\vspace{1mm}

\noindent\textbf{\underline{\clk{}}.} This is a QoS-aware query distribution scheme inspired by Clockwork~\cite{gujarati2020serving}. It has a focus on latency consolidation, but we have used its central controller for comparison with \sol{}. This scheme monitors all hardware availability timings and accurately predicts query latency. A query is guaranteed to be served within its latency target unless none of the instances can meet the QoS target. Each instance maintains an individual FCFS query queue. There is a central controller to keep track of all hardware and queue timings and send new queries to the instance queues. \vspace{1mm}

\noindent\textbf{\underline{ORCL}.} Oracle scheme is a practically infeasible scheme only for reference to understand the limits of performance. Oracle always has higher throughput than other schemes because it knows the future query arrival patterns. It creates a sequence of queries according to batch size distribution and sorts all queries by the batch size. Whenever a base instance is available, it serves the next largest query. For auxiliary instances, it is the next smallest query. There is no wait time for queries, and queries never run on instances that violate QoS. The throughput on serving the query sequence is recorded, and among all possible heterogeneous configurations, the largest throughput is used as the Oracle throughput. 

Recall that \textbf{\underline{\sol{}}} determines the near-optimal heterogeneous configuration using its upper-bound-assisted method without any online evaluation - unlike other competing schemes. \textbf{\underline{\solplus{}}} is an online variation of \sol{} for optimal configuration, but the number of online evaluations is limited because it is guided by \sol{}'s upper bound method, as our evaluation discusses next. 

\section{\sol{}: Experimental Evaluation}
\label{sec:evaluation}
\begin{figure}
    \centering
    \includegraphics[scale=0.5]{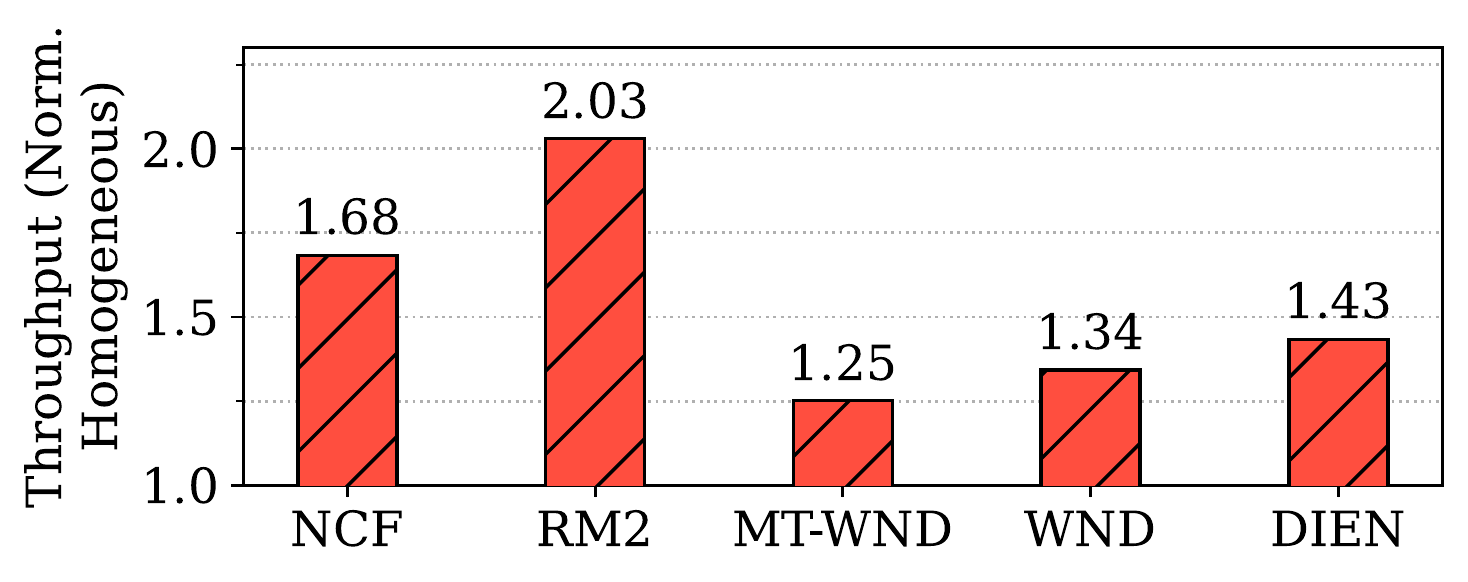}
    % \vspace{0.1cm}
    \hrule
    \vspace{-0.4cm}        
    \caption{\sol{} yields higher throughput compared to the homogeneous configuration.}
    \vspace{-0.3cm}
    \label{fig:eval_0_main}
\end{figure}
First, we quantitatively evaluate \sol's performance compared to other competing schemes and state-of-the-art approaches. We explain why \sol{} works effectively, and what the key contributing factors toward its superiority over existing methods are. Finally, we demonstrate that \sol{} performs effectively even under varying constraints and parameters (e.g., cost budget, QoS, and batch size distribution). 

\subsection{Comparison with Best Homogeneous}
First, our experimental results (Fig.~\ref{fig:eval_0_main}) confirm that \sol{} provides significant allowable throughput improvement compared to the most competitive base instance homogeneous configuration under the same QoS constraints and budget (Sec.~\ref{sec:methodology}). 
The number of instances in the homogeneous pool is determined by the maximum number of nodes that can fit within the cost budget, this is the optimal homogeneous resource configuration. Note that the cost budget is not a multiple of \texttt{g4dn.xlarge} price. To compensate for the wasted budget, we scale up the homogeneous throughput proportionally. \textit{But, to evaluate conservatively, we allow the budget slack of \sol{} to be wasted. Despite this, \sol{} is still able to provide up to $2\times$ throughput (i.e., RM2 model) and more than $1.25\times$ in all cases. }

\subsection{Comparison with State-of-the-Art}

%Techniques

\begin{figure}
    \centering
    \includegraphics[scale=0.49]{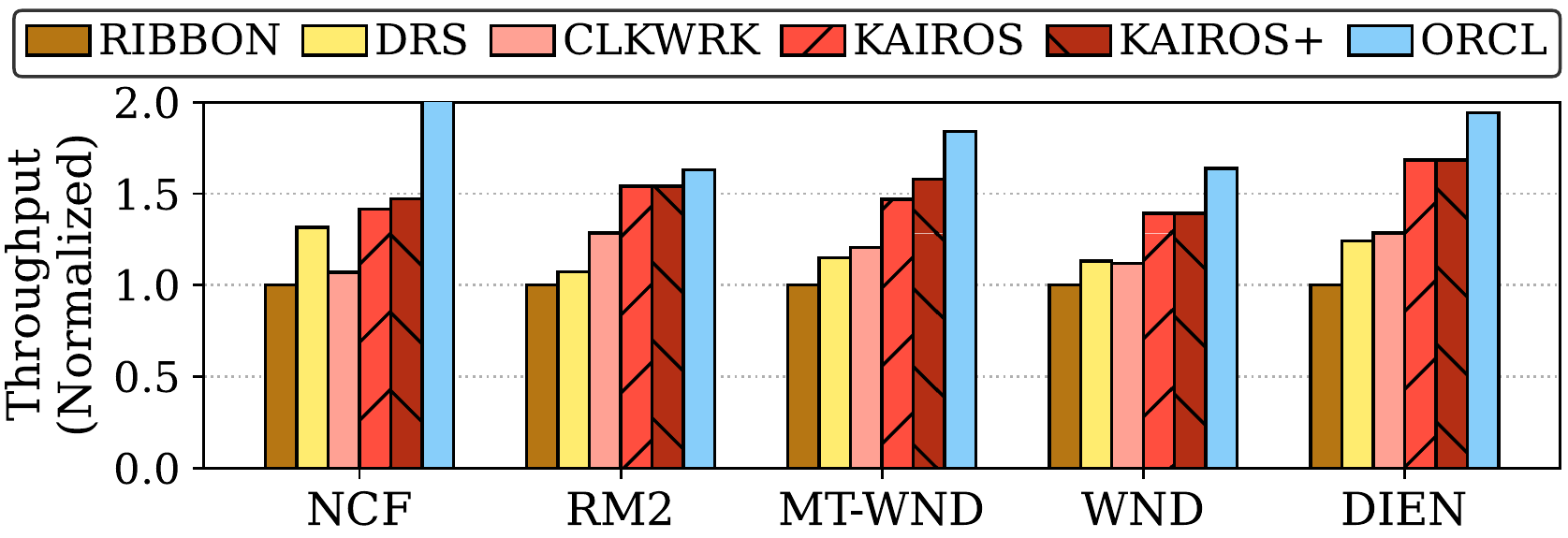}
    \vspace{0.1cm}
    \hrule
    \vspace{-0.4cm}          
    \caption{Throughput comparison of \sol{} and \solplus{} against other competing schemes.}
    \vspace{-0.3cm}
    \label{fig:eval_4}
\end{figure}

Next, our results show that \sol{} outperforms competing schemes of \ribbon{}, DRS, and \clk{} in Fig.~\ref{fig:eval_4}. Recall that the competing techniques \drs{} and \clk{} only focus on distributing the queries to a set of instances following a certain strategy. They do not observe that one can build a heterogeneous configuration that is more performant than a homogeneous configuration under a cost budget. To provide these techniques an advantage, we provide each scheme with the best heterogeneous configuration obtained via Oracle search and compare the throughput against \sol{}. Fig.~\ref{fig:eval_4} shows the allowable throughput for different schemes.

We make several observations. First, as expected, both DRS and \clk{} outperform the \ribbon{} scheme because \ribbon{} uses a simple query distribution mechanism. Second, \sol{} significantly outperforms all existing techniques and is close to the Oracle scheme. For all models, \sol{} can provide around $1.5\times$ the throughput of \ribbon{}. \sol{}'s performance comes from its superior query distribution mechanism (Sec.~\ref{sec:design}) which focuses on exploiting the heterogeneity for maximum availability cycles on all resources combined, while carefully avoiding QoS violations. Note that DRS is ultimately limited by its threshold-based mechanism and misses the opportunity of utilizing different instance types for smaller queries. \clk{} actively schedules queries on instances that do not violate QoS, thus it helps avoid more unnecessary QoS violations than \ribbon{}. However, unlike \sol{}, it does not optimize on heterogeneous instances.

The most attractive aspect of \sol's superior performance and design is that \sol{} does not experimentally evaluate heterogeneous configurations online -- unlike other methods. It uses its unique approximation method to determine a good heterogeneous configuration in one shot. \textit{For our evaluation, we provided additional advantages to all competing schemes by allowing them to use optimal heterogeneous configuration -- determined offline.} Even under this conservative evaluation, \sol{} outperforms the DRS and \clk{} schemes by up to 44\% (Fig.~\ref{fig:eval_4}).

Lastly, \solplus{} performs slightly better than \sol{}. This is expected because \solplus{} employs a pruning-based online approach to find the optimal heterogeneous configuration. Nevertheless, \sol{} still provides approximately the same throughput as \solplus{} without any online evaluation. Next, we discuss the impact of online configuration searches.

\subsection{Online Optimal Config. Exploration}
\label{sec:eval_online}

\begin{figure}
    \centering
    \includegraphics[scale=0.5]{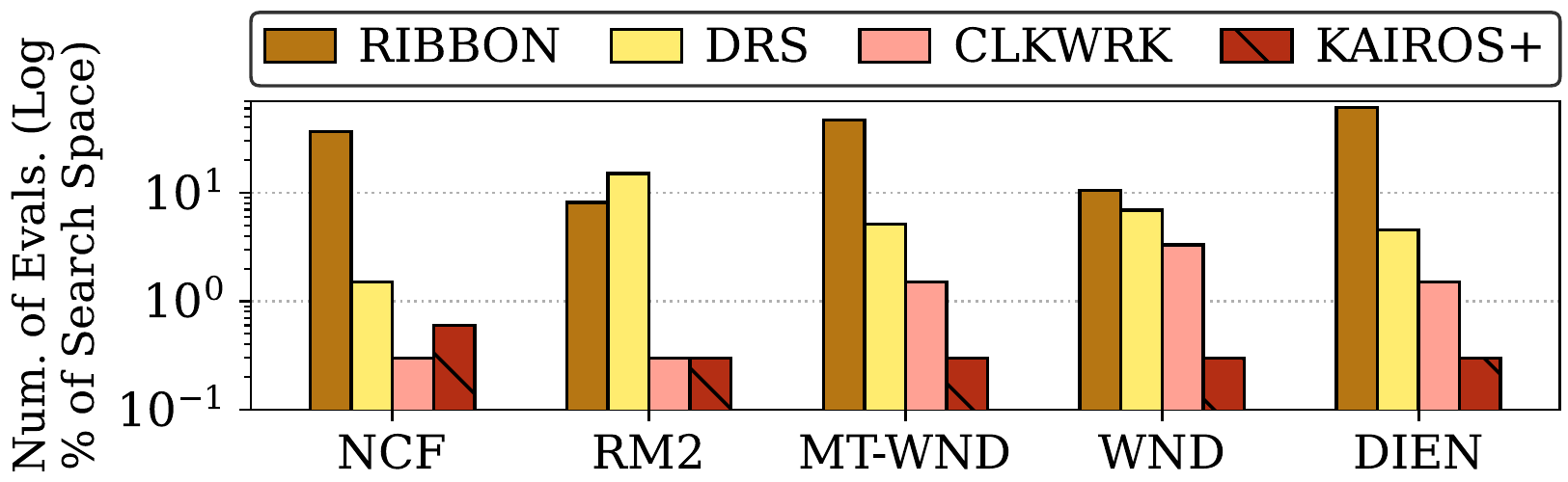}
    \vspace{0.1cm}
    \hrule
    \vspace{-0.4cm}         
    \caption{\solplus{} has low evaluation overhead.}
    \vspace{-0.3cm}    
    \label{fig:eval_5}
\end{figure}

Among all competing schemes, only \sol{} and Oracle do not require online evaluation, others need to evaluate configurations online to find the optimal one for their query distribution mechanism. One may ask: how much overhead does \sol{} save from avoiding online evaluation? In our previous result (Fig.~\ref{fig:eval_4}), the impact of online evaluation is not included. Fig.~\ref{fig:eval_5} shows the number of evaluations each online technique requires to find its optimal configuration (on log scale). We observe that \ribbon{} and DRS often need to evaluate 5\%-30\% of the search space to reach the optimal configuration, and hence, the maximum throughput they can achieve is \textit{delayed} by the length of this exploration period. 

For a fair comparison, in Fig.~\ref{fig:eval_5}, we augment all competing techniques with the same online configuration exploration algorithm as \solplus{}. We note that \solplus{} consistently evaluates less than 1\% of the search space for all models, outperforming competing schemes despite using the same search algorithm. Similar trends are observed, even with other non-\solplus{} online exploration algorithms such as genetic algorithm, simulated annealing, etc. 

\begin{figure}
    \centering
    \includegraphics[scale=0.5]{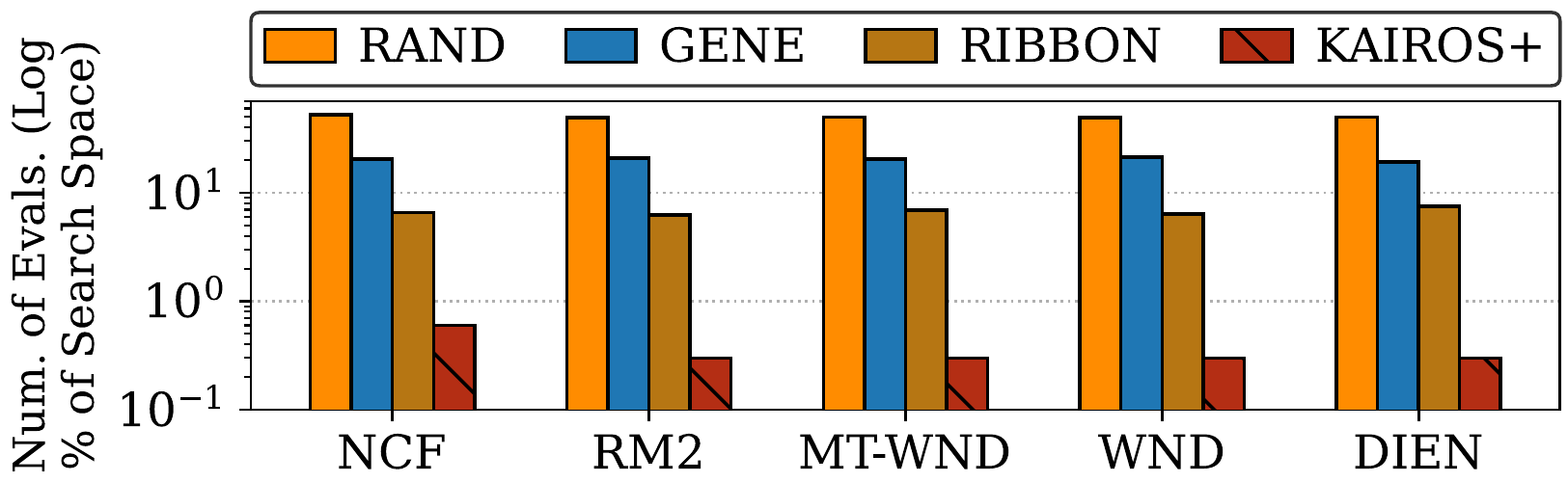}
    \vspace{0.1cm}
    \hrule
    \vspace{-0.4cm}         
    \caption{\solplus{} versus competing search algorithms.}
    \vspace{-0.3cm}
    \label{fig:eval_1}
\end{figure}

A related inquiry is: is calculating the throughput upper bound and leveraging it toward online exploration in \solplus{} useful? Will any other alternative search methods such as \ribbon{} quickly find the same optimal configuration that \solplus{} finds? We use random search (RAND), genetic algorithm~\cite{wortmann2016black} (GENE), and \ribbon{}'s Bayesian Optimization as competing search algorithms and evaluate the number of evaluations compared to the \solplus{} technique. We purposely provide these competing algorithms with the same sub-configuration pruning mechanism as \solplus{} in Algorithm~\ref{algo:greedy} to save some iterations. But, as Fig.~\ref{fig:eval_1} shows, competing methods still require significantly more online evaluations than \solplus{}. These results essentially decouple the effects of \sol{}'s two components of query distribution and search space evaluation, showing that they both are critical for quickly achieving a configuration that performs well. The advantages of \sol{} as a technique that can quickly find a good configuration are further emphasized.

\begin{figure}
    \centering
    \includegraphics[scale=0.5]{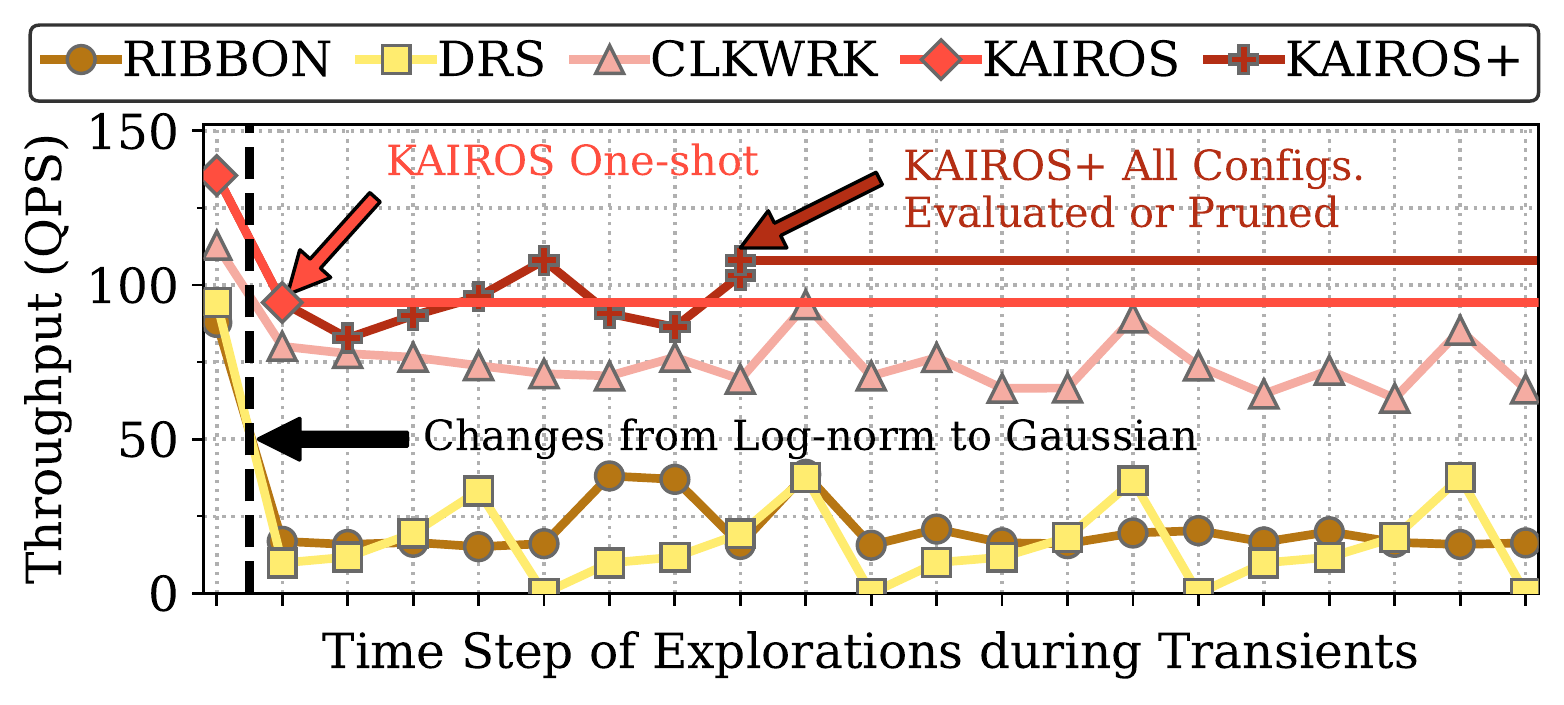}
    % \vspace{0.1cm}
    \hrule
    \vspace{-0.4cm}     
    \caption{When query size probability distribution changes, the throughput of evaluated configurations.}
    \vspace{-0.3cm}
    \label{fig:eval_10}
\end{figure}
\subsection{Timely Reaction to Load Changes}
\label{sec:eval_transient}

Previous results demonstrate that, besides a higher throughput, \sol{}'s major benefit is from eliminating the time overhead to find a promising configuration. Fig.~\ref{fig:eval_10} provides further experimental demonstration to substantiate this. When the query load or its distribution changes, the optimal configuration changes. In Fig.~\ref{fig:eval_10}, the query-size distribution changes from Log-normal to Gaussian (at the vertical dashed line) for a sample model (RM2), and the first 20 evaluated configurations of transient response are shown. All schemes respond to this change and restart the search process.

In Fig.~\ref{fig:eval_10}, \sol{} reaches a near-optimal configuration in one shot. Without online evaluation, its throughput is $2\times$ more than that of \ribbon{} and DRS. \clk{} is the most competitive scheme, but it uses 9 evaluations to reach the same throughput as \sol{}. To find the optimal configuration, \solplus{} performs upper-bound-assisted online search: by its $8^{th}$ evaluation, all possible configurations in the search space have been either evaluated or pruned by upper bound, and as expected, its final throughput is slightly higher than \sol{}. The throughput of \sol{} and \solplus{} are within 15\% of the Oracle (not shown for better figure readability).

\subsection{Source of \sol{}'s Effectiveness}
\label{sec:eval_why}

\begin{figure}
    \centering
    \includegraphics[scale=0.44]{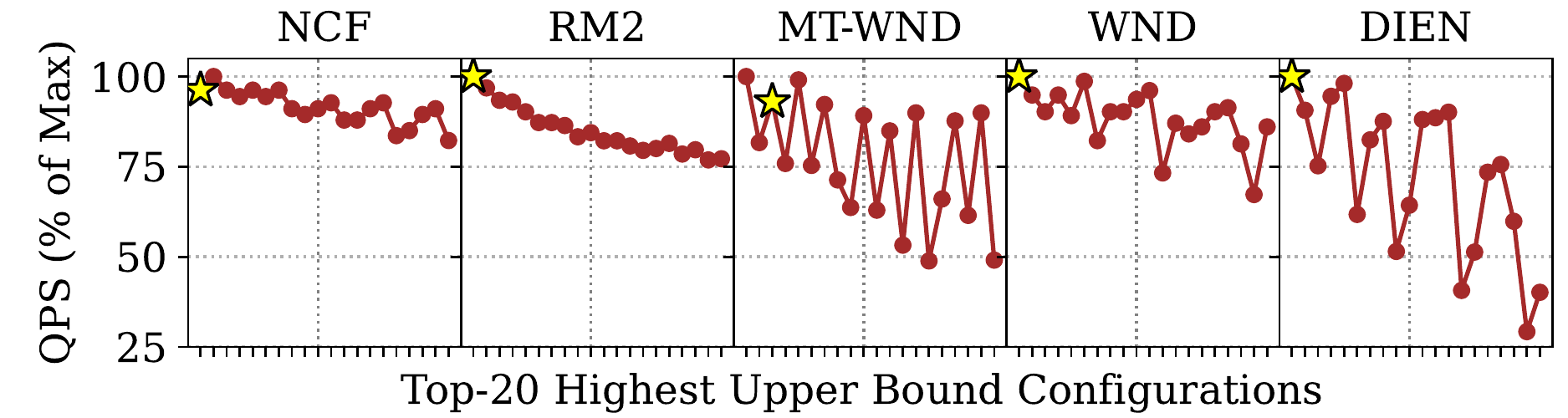}
    \vspace{0.1cm}
    \hrule
    \vspace{-0.4cm}         \caption{Actual throughput of the configs. sorted by upper bounds. \sol{}'s selected configuration is marked by star.}
    \vspace{-0.3cm}
    \label{fig:eval_2}
\end{figure}

\sol{}'s effectiveness comes from its upper-bound method that quickly finds a near-optimal heterogeneous configuration. Fig.~\ref{fig:eval_2} explains why it can find such configurations. In this figure, the top-20 highest upper bound configurations are shown with their actual throughput in red. The star corresponds to the configuration that \sol{} picks after applying its similarity-based criteria to choose the most promising one from the top 10 configurations. 
We make two key observations. Firstly, the actual optimal configuration is always among the top 10 candidates. Secondly, although the actual throughput does not follow the upper bounds in strict monotonic order (i.e., a higher upper bound always means higher throughput), they still follow the same trend, indicating the optimal configuration is among the highest upper bound ones. 
This explains why \sol{} can find a near-optimal configuration without online evaluations.

\begin{figure}
    \centering
    \includegraphics[scale=0.5]{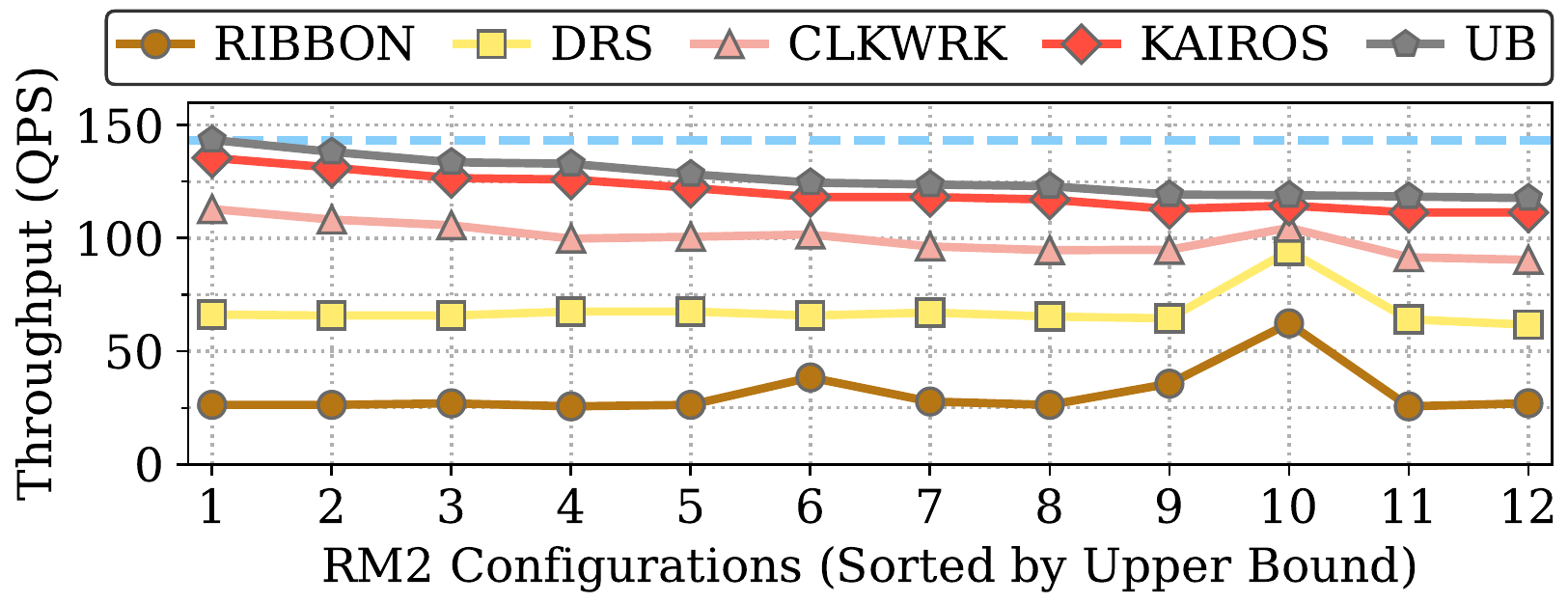}
    % \vspace{0.1cm}
    \hrule
    \vspace{-0.4cm}         
    \caption{Impact of changing query distribution scheme. The dashed horizontal line indicates Oracle.}
    \vspace{-0.3cm}
    \label{fig:eval_3}
\end{figure}

To further demonstrate the source of effectiveness of \sol{}, in Fig.~\ref{fig:eval_3}, we pick the RM2 model and plot the experimental throughput for \sol{}'s top upper bound configurations when changing the query distribution scheme to \ribbon{}, DRS, and \clk{}. This setting is chosen to better understand how the query distribution mechanism and heterogeneous configuration search by \sol{} are co-designed, and replacing the query distribution mechanism with any other would result in a worse performance of configuration search.  
% From this figure, 
We make three observations: (i) The upper bound (UB) is lower than but close to the Oracle throughput (dashed), indicating that the upper bound is relatively tight and meaningful to be used in practice. (ii) The calculated (UB) and \sol{}'s experimentally observed throughput are close to each other and follow the same trend. This substantiates the design decision to approximate real throughput using the upper bound. (iii) \sol{}'s query-distribution mechanism is a key source of its effectiveness. If \sol{} only chooses a configuration based on the upper bound without employing its query-distribution mechanism, the actual throughput will yield far from expectation (loose bound) -- underscoring the effectiveness of \sol{}'s query-distribution mechanism. 

\subsection{Parameter Robustness Evaluation}
\label{sec:eval_robust}

\begin{figure}[t]
    \centering
    \includegraphics[scale=0.43]{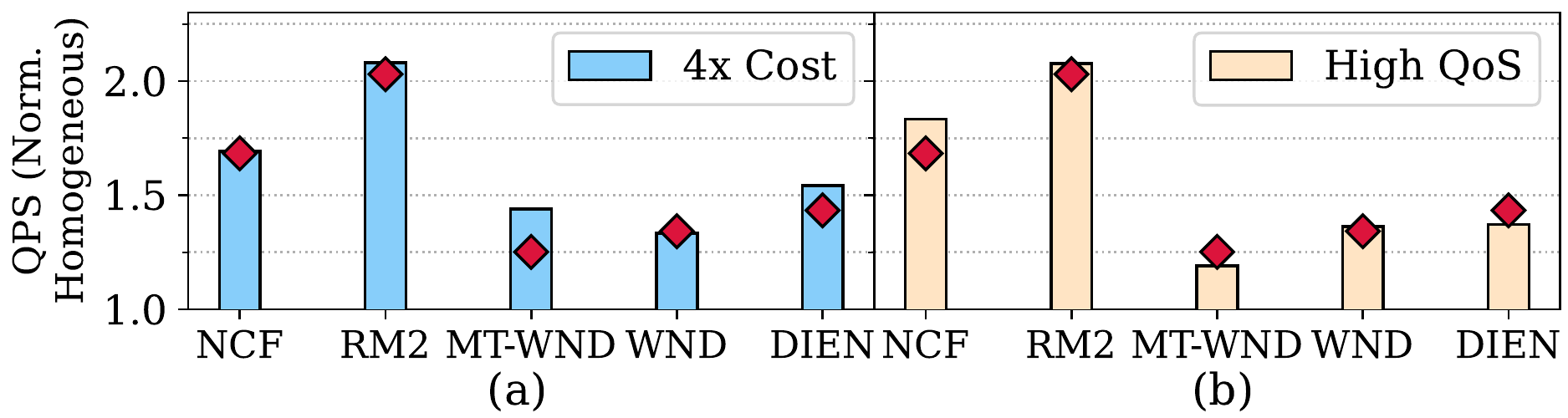}
    \vspace{0.1cm}
    \hrule
    \vspace{-0.4cm}         
    \caption{\sol{} when: (a) budget is scaled by 4x; (b) using a higher QoS target. Red dot: Fig.~\ref{fig:eval_0_main} results.}
    \vspace{-0.3cm}
    \label{fig:eval_7}
\end{figure}

Finally, we evaluate \sol{}'s robustness against different parameters. Fig.~\ref{fig:eval_7}(a) shows that \sol{}'s heterogeneity approach offers a substantial improvement over homogeneous configurations when the cost budget scales. Notice that non-\sol{} schemes would struggle more in finding a good configuration as the search space is increased by $4\times$. Similarly, in Fig.~\ref{fig:eval_7}(b), when the QoS targets are set 20\% higher, \sol{} continues to offer similar improvements as before (Fig.~\ref{fig:eval_0_main} results are marked by the red dot).

\begin{figure}[t]
    \centering
    \includegraphics[scale=0.43]{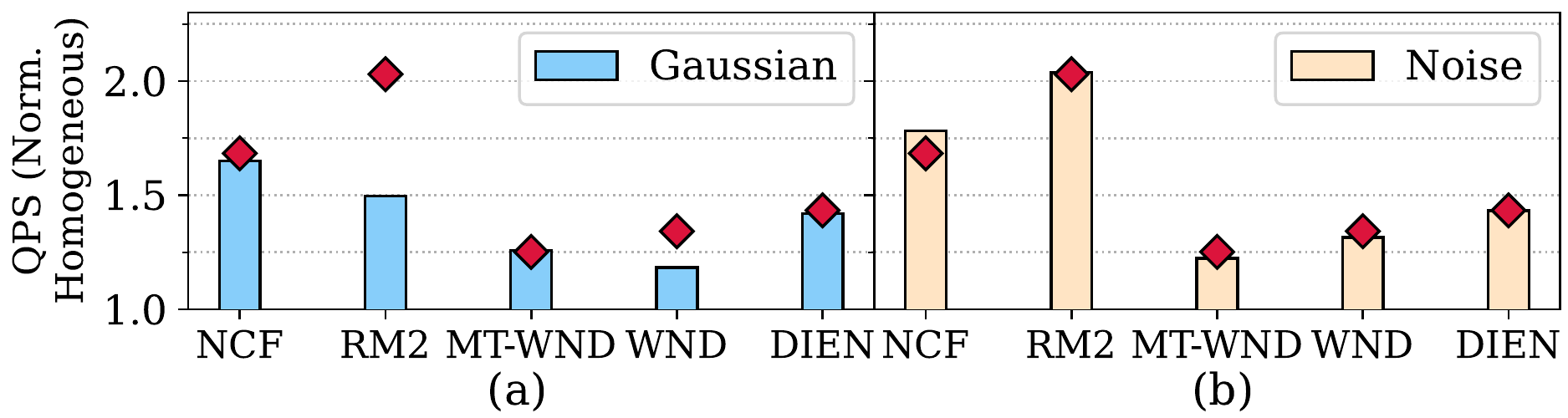}
    \vspace{0.1cm}
    \hrule
    \vspace{-0.4cm} 
    \caption{\sol{}'s effectiveness when: (a) batch size follows Gaussian distribution; (b) query latency noise presents.}
    \vspace{-0.3cm}
    \label{fig:eval_9}
\end{figure}

Recall that the query batch size patterns may change over time, and \sol{} can quickly respond to the new patterns without online evaluations (Sec.~\ref{sec:eval_transient}). In Fig.~\ref{fig:eval_9}(a), we show that \sol{} still yields significant benefits over homogeneous serving on Gaussian distributed batch sizes. Note that the Oracle could also yield lower improvements -- hence, the relative improvement compared to the previous distribution (red dots) also decreases for some models. Lastly, since the query distribution scheme of \sol{} makes a realistic assumption that the inference latency can be accurately predicted, we also show \sol{}'s effectiveness when this assumption is relaxed. In Fig.~\ref{fig:eval_9}(b), we intentionally inject an additive Gaussian white noise with 5\% variance in latency prediction to emulate performance variability in the cloud~\cite{ericson2017analysis}. Our results suggest that \sol{} is not sensitive to such noise that is common due to interference or transient hardware degradation, and continues to offer similar improvements.

\section{Conclusion}
\label{sec:relat}
\sol{} demonstrates that a mixture of heterogeneous compute instances can be effectively utilized to maximize the inference query throughput under QoS and cost constraints. \sol{}'s upper-bound-based method eliminates the need for online configuration exploration, and \sol{} intelligently distributes queries among heterogeneous instances. Our evaluation shows that \sol{} significantly outperforms state-of-the art techniques. We expect that \sol{}'s two-pronged approach could potentially find interesting applicability in other computer system optimization problems.

\begin{acks}
This material is based upon work supported by the Assistant Secretary of Defense for Research and Engineering under Air Force Contract No. FA8702-15-D-0001, and United States Air Force Research Laboratory Cooperative Agreement Number FA8750-19-2-1000. Any opinions, findings, conclusions or recommendations expressed in this material are those of the author(s) and do not necessarily reflect the views of the Assistant Secretary of Defense for Research and Engineering, or the United States Air Force. The U.S. Government is authorized to reproduce and distribute reprints for Government purposes notwithstanding any copyright notation herein.
\end{acks}

\bibliographystyle{unsrtnat}
\balance
\bibliography{refs}

\end{document}